\documentclass[10pt,english,aps,pra,superscriptaddress,amsmath,amssymb,preprintnumbers,twocolumn,showpacs,showkeys]{revtex4-1}
\usepackage[utf8]{inputenc}
\usepackage{color}
\usepackage{babel}
\usepackage{amsmath}
\usepackage{amssymb}
\usepackage{graphicx}
\usepackage{esint}
\usepackage[unicode=true,pdfusetitle,
 bookmarks=true,bookmarksnumbered=false,bookmarksopen=false,
 breaklinks=false,pdfborder={0 0 0},backref=false,colorlinks=true]
 {hyperref}
\hypersetup{
 linkcolor=blue,citecolor=blue,urlcolor=blue}

\makeatletter
 
 \@ifundefined{textcolor}{}
 {%
   \definecolor{BLACK}{gray}{0}
   \definecolor{WHITE}{gray}{1}
   \definecolor{RED}{rgb}{1,0,0}
   \definecolor{GREEN}{rgb}{0,1,0}
   \definecolor{BLUE}{rgb}{0,0,1}
   \definecolor{CYAN}{cmyk}{1,0,0,0}
   \definecolor{MAGENTA}{cmyk}{0,1,0,0}
   \definecolor{YELLOW}{cmyk}{0,0,1,0}
 }

\usepackage{epsfig}
\usepackage{enumerate}

\makeatother

\begin{document}

\title{Quantum correlations of identical particles subject to classical
environmental noise}

\author{Andrea~Beggi}

\email{andrea.beggi@unimore.it}

\selectlanguage{english}%

\affiliation{Dipartimento di Scienze Fisiche, Informatiche e Matematiche, Università
di Modena e Reggio Emilia \\
 Via Campi 213/A, I-41125, Modena, Italy}

\author{Fabrizio~Buscemi }

\email{fabrizio.buscemi@unimore.it}

\selectlanguage{english}%

\affiliation{Dipartimento di Scienze Fisiche, Informatiche e Matematiche, Università
di Modena e Reggio Emilia \\
 Via Campi 213/A, I-41125, Modena, Italy}

\author{Paolo~Bordone}

\affiliation{Dipartimento di Scienze Fisiche, Informatiche e Matematiche, Università
di Modena e Reggio Emilia and \\
 Centro S3, CNR-Istituto di Nanoscienze, Via Campi 213/A, I-41125,
Modena, Italy}
\begin{abstract}
In this work we propose a measure for the \emph{quantum discord }of
indistinguishable particles, based on the definition of \emph{entanglement
of particles} given in\emph{ } [H. M. Wiseman \emph{et al.}, \href{http://www.dx.doi.org/10.1103/PhysRevLett.91.097902}{Phis. Rev. Lett {\bf 91}, 097902 (2003)}].
This \emph{discord of particles} is then used to evaluate the quantum
correlations in a system of two identical bosons (fermions), where
the particles perform a quantum random walk described by the Hubbard
Hamiltonian in a 1D lattice. The dynamics of the particles is either
unperturbed or subject to a classical environmental noise – such as
random telegraph, pink or brown noise. The observed results are consistent
with those for the \emph{entanglement of particles}, and we observe
that on-site interaction between particles have an important protective
effect on correlations against the decoherence of the system.
\end{abstract}



\maketitle

\section{Introduction}

Since its first introduction\cite{Einstein1935,*Schroedinger1935,*Schroedinger1935b,*Schroedinger1935c},
quantum \emph{entanglement} has become one of the most intriguing
and characteristic traits of quantum mechanics. In the last two decades,
it has turned into a fundamental resource for quantum information
theory and quantum computing\cite{nielsen2010quantum,benenti2007principles},
since it can be used to implement protocols and calculations that
would be impossible in a classical context. However, recent developments
in the field of quantum information showed that the entanglement is
not able to capture all the quantum correlations contained in a system,
and therefore some separable (i.e. not entangled) states still possess
a certain amount of correlations that can be used to perform non-classical
computing tasks\cite{Biham2004,Lanyon2008}. Thus, in order to take
into account these correlations, many quantifiers were proposed in
recent years\cite{Modi2010,Modi2012}, among which the most used is
certainly \emph{quantum discord}\cite{Ollivier2001,Henderson2001}.
Quantum discord represents a measure of the (quantum) correlations
of a system that are destroyed by a measurement on a subparty of the
system, and in general it does not coincide with entanglement, nor
entanglement is necessarily contained within discord\cite{Luo2008a}.

Although there are many possible quantifiers and witnesses for entanglement
in multipartite systems\cite{Guhne2009}, the entanglement in bipartite
systems of distinguishable particles has a well defined formulation\cite{Bennett1996,Wooters1998,Horodecki2009}.
The same consideration does not hold for systems of identical particles,
where many definitions have been introduced in recent years, raising
debates about the reliability of the proposed quantifiers\citep{Schliemann2001, *Eckert2002, Zanardi2002, Buscemi2006, *Buscemi2007, Ghirardi2002, *Ghirardi2004, Benatti2014, Wiseman2003, Dowling2006, Sasaki2011, Iemini2013, Iemini_PhysRevA.87.022327, *Iemini_PhysRevA.89.032324, *Reusch_PhysRevA.91.042324}.
The main difficulties appearing in the estimation of entanglement
for indistinguishable particles arise from the exchange symmetry,
which requires the (anti-)symmetrization of the wavefunction: as a
consequence, pure states cannot be factorized anymore, even if the
particles are not entangled. This requires the introduction of a specific
criterion that can distinguish between the non-separability due to
genuine entanglement and the ``spurious'' correlations due to exchange
symmetry, which cannot be used to violate Bell’s inequality and therefore
are not a resource for quantum-information processing\citep{Ghirardi2002, *Ghirardi2004}.

Among the different proposed approaches for calculating the entanglement
of identical particles, the Wiseman and Vaccaro criterion\cite{Wiseman2003}
can overcome some problems shown by other methods \cite{Dowling2006}.
It has been recently extended to multipartite systems\cite{Buscemi2011}
and it was applied for studying the correlations in quantum walks
of identical non-interacting particles\cite{Benedetti2012_QWferbos}
and the entanglement between sites in Hubbard spin chains\cite{Mazza_1367-2630-17-1-013015},
where it also acts as an order parameter which is able to capture
quantum phase transitions\cite{Iemini_PhysRevB.92.075423}. Indeed,
it was also suggested that this kind of entanglement could be measurable
in many experimental scenarios \cite{Dowling2006}, thus making it
an interesting quantifier for quantum correlations.

At present, however, the evaluation of quantum discord of identical
particles is still an almost unexplored topic, with few exceptions
\cite{Iemini2013,Wang2010}. It is well known, however, that quantum
discord possesses some features that make it more promising than entanglement
in accounting for correlations - such as e.g. a higher robustness
under decoherence, the general absence of sudden death phenomena and
the ability to identify quantum phase transitions that are missed
by entanglement \cite{Modi2012}.

In the last years, the interest in entanglement among identical particles
is increased, since it is crucial for the understanding of many physical
phenomena which involve highly correlated indistinguishable subsystems,
such as photons in nonlinear waveguides\cite{Bromberg2009,Peruzzo2010,Lahini2012},
ultracold atoms trapped in optical lattices\cite{RevModPhys.80.885,RevModPhys.80.1215,fukuhara2013,fukuhara2013b}
or electrons in solid state systems\cite{PhysRevB.63.085311,ZanardiWang2002(JPA)}.
These systems constitute a possible prototype for implementing quantum
computing devices, and indeed some experimental realizations of photonic
chips have been achieved recently\cite{Broome2013,Spring2013,Tillmann2013},
whose architecture rely on photonic quantum walks. The most basic
Hamiltonian that can describe a continuous-time Quantum walk (QW)\cite{Kempe2003_QW,Venegas2012_QW}
in those physical systems is the Hubbard Hamiltonian\citep{essler,*montorsi1992hubbard,*lieb2003one},
which has also been used as a benchmark for entanglement criteria
of indistinguishable particles\cite{Zanardi2002,Dowling2006}. Besides
this, QWs of identical particles are of peculiar interest since the
indistinguishability of the walkers is responsible for the building-up
of genuinely quantum correlations, even in the absence of interactions
between particles (e.g. photons) \cite{Franson2013,Benedetti2012_QWferbos,Mayer2011,SamuelssonButtiker_ABEffect}.

In real physical systems QWs are subject to environmental noise, whose
effect is to destroy quantum correlations among the walkers\cite{Marzolino2013}.
Therefore, the study of decoherence is of vital importance for the
realization of devices that are able to implement robust quantum information
protocols, in order to preserve correlations against the action of
the environment. A possible model to represent an external noise source
appearing in many nanodevices is a random bistable fluctuator with
switching frequency $f_{0}$, which produces the so-called \emph{random
telegraph noise} (RTN)\cite{Fujisawa2000,Kurdak1997}. A collection
of bistable fluctuators with different switching rates $f$ can also
be used to model\cite{Benedetti2013} \emph{colored noises}, such
as $1/f$ (``pink'') or $1/f^{2}$ (``brown'') noises, which are
very common in solid-state physics\citep{kogan2008electronic,*Weissman1988_REV,*Vandamme1994,*milotti20021}.
There are many works in quantum information literature focused on
RTN\citep{Bordone2012, Buscemi2013, De2011, Franco2012, *Wold2012, *Mazzola2011, *Zhong2010}
or $1/f$ noises\citep{De2011, Paladino2014, Kakuyanagi2007, *Yoshihara2006, Paladino2002, *Paladino2005, *Paladino2010, *Paladino2011},
but few studies are available for $1/f^{2}$ noise \cite{Martinis2003,Anton2012,Benedetti2013,Benedetti2013_ICNF}.
The exploration of these kind of noises is of utmost importance for
practical applications, since the system can exhibit peculiar phenomena
(like sudden death and revival of correlations, or memory effects)
as a function of the noise spectrum\citep{Yu2009, *Yu2004, *Bellomo2007},
even for classical noise sources\citep{Leggio2015, *xu2013, *zhou2010, *LoFranco2012, Bordone2012, Benedetti2012, Benedetti2013}
such as RTN fluctuators%
\footnote{We refer to a noise as \emph{classical }when the system is coupled
with the environment but the environment is unaffected by the system,
i.e. no back-action-induced correlations can be transferred from the
system to the environment. Classical environments, however, can have
``effective'' back-action on a physical system, as recently shown\citep{Leggio2015, *xu2013, *zhou2010, *LoFranco2012, Bordone2012, Benedetti2012, Benedetti2013},
thus determining a revival of quantum correlations.%
}. 

For what concerns QWs, many studies are devoted to noise in discrete-time
quantum walks, but for continuous-time quantum walks there is a limited
amount of studies\citep{Kendon2007_noiserev, *hines2008}, which are
concerning e.g. static noise \cite{Yin2008}, RTN \cite{Bordone2012,Prokofev2006}
or other mechanisms of decoherence, such as phonon thermal baths (see
\cite{Nizama2014} and cited references), unitary noise\cite{Shapira2003},
measurements\cite{Fedichkin2006} and lattice defects\cite{Romanelli2005}.
Among those, very few are focused on the time-evolution of quantum
correlations. It seems, therefore, of interest to realize a general
overview of the effects of noise on continuous-time QWs correlations,
keeping also into account the effects due to indistinguishability
of the walkers.

The aim of this paper is double: first of all, we perform an extensive
study on quantum correlations in continuous-time QWs of identical
particles (fermions and bosons), whose dynamics is either unperturbed
or subject to a classical noisy environment – namely a single RTN
fluctuator or a collection of bistable fluctuators mimicking colored
noises ($1/f$ and $1/f^{2}$), in order to quantify the role of classical
noise in the decoherence of the system. Secondly, in order to fully
characterize quantum correlations, we introduce a measure for the
quantum discord of identical particles, and we confront it with the
corresponding value of entanglement, to prove that it is a good quantifier
of quantum correlations. The dynamics of quantum correlations are
then explored as a function of many physical parameters, such as the
strength of interactions among the walkers, the number of open decohering
channels and the frequency of the noise sources.

The paper is organized as follows. In Sect. \ref{sec:Bipartite-entanglement-and-discord}
we first review the entanglement criteria for identical particles
introduced in the literature and their known problems; after this,
we illustrate the concept of \emph{entanglement of particles} as introduced
by Wiseman and Vaccaro\cite{Wiseman2003}, then we extend this approach
in order to introduce the \emph{quantum discord of particles}. Then,
in Sect. \ref{sec:Physical-models} we introduce and characterize
the physical models that describe the quantum walks of our fermionic
and bosonic particles – which are both based upon the Hubbard Hamiltonian
– and the mechanisms that we exploit to generate RTN and colored noises.
Sections \ref{sec:Numerical-simulations:-Bose-Hubb}-\ref{sec:Numerical-simulations:-Fermi-Hubb}
are devoted to the discussion of the results of the numerical simulations:
the first one is concerning bosons, the second one is about fermions.
Finally, in Sect. \ref{sec:Conclusions} some conclusions are drawn,
and some further perspectives of research are suggested.

\section{Bipartite entanglement and discord of identical particles\label{sec:Bipartite-entanglement-and-discord}}

In this section, we shortly review the criteria used in the literature
to estimate the entanglement in bipartite systems of two indistinguishable
particles, then we illustrate the quantifier that we adopted in our
work, which we extend to the evaluation of quantum discord. Specifically,
our approach relies on the concept of \emph{entanglement of particles},
introduced by Wiseman and Vaccaro~\cite{Wiseman2003,Dowling2006}
in order to give an operational definition of the entanglement of
two identical particles which does not violate the superselection
rule of the local particle number. 

One of the first proposed quantifiers for the entanglement of identical
particles was the Schliemann criterion\citep{Schliemann2001, *Eckert2002},
which is based upon the Slater decomposition of the (anti-)symmetrized
state and evaluates the quantum correlations as a function of the
minimum Slater rank of the considered state (over all its possible
decompositions). Unfortunately, this approach behaves incorrectly
under local and nonlocal mode transformations when the entanglement
is evaluated between the modes of a system\cite{Gittings2002}. The
same problem affects the entanglement measures proposed in Refs.~\cite{Paskauskas2001,Li2001}.
To overcome these limits, another criterion was proposed by Zanardi
\cite{Zanardi2002}: in his approach, the space of the modes of the
systems is mapped into qubit states, and the entanglement between
particles is evaluated as the entanglement between modes. It has been
recently demonstrated\cite{Benatti2014} that this criterion is equivalent
to those that identify pure separable states as states where both
parties possess a complete set of defined properties\citep{Ghirardi2002, *Ghirardi2004}.
The approach used by Zanardi, however, can lead to the overestimation
of entanglement due to violation of the local number of particles
superselection rule, and this is the reason why it has been generalized
by Wiseman and Vaccaro\cite{Wiseman2003}. In their proposal, the
so-called \emph{entanglement of particles} is represented by the maximum
amount of nonclassical correlations that can be extracted (``accessible
entanglement'') from the system with local operations, and then can
be encoded into conventional \emph{quantum registers} (i.e. a set
of distinguishable qubits). Within this definition, the problems of
the aforementioned criteria can be overcome. Other approaches investigated
in the literature define entanglement via non-classical correlations
between subsets of observables\citep{Zanardi2004, *Barnum2004, *Benatti2012, *Benatti2014b},
but do not possess the same easiness of computability of this criterion. 

To go into further detail, the Wiseman and Vaccaro criterion addresses
the non-classical correlations in the form of entanglement between
two distant parties (namely Alice and Bob) of a quantum system in
the mixed state $\rho$, each accessing a given set of modes. Any
subsystem is assumed to have a standard quantum register, that is
a set of distinguishable qubits, in addition to the indistinguishable
particles described by $\rho$. The \emph{entanglement of particles}
$E_{P}$ is given by the maximum amount of entanglement that Alice
and Bob can produce between their standard quantum registers by means
of local operations on the modes that they have access to. For a two-particle
system, $E_{P}$ can simply be expressed as 
\begin{equation}
E_{P}=P_{1,1}\mathcal{E}(\rho_{1,1})\label{eq:entapart}
\end{equation}
where $\rho_{1,1}=\Pi_{1,1}\rho\Pi_{1,1}$ is the state obtained from
$\rho$ by means of the projectors $\Pi_{1,1}$ onto the state having
one particle in each subsystem. $P_{1,1}$ denotes the probability
of finding 1 in measurements of the local number of particles by both
Alice and Bob, and $\mathcal{E}$ represents a bipartite standard
entanglement measure estimating the degree of non classical-correlation
between the quantum registers of distinguishable qubits controlled
by Alice and Bob. Specifically, here the latter is evaluated in terms
of the entanglement of formation~\cite{Wooters1998}: 
\begin{equation}
\mathcal{E}=h\left(\frac{1+\sqrt{1-C^{2}}}{2}\right),
\end{equation}
where $\quad h(x)=-x\ln_{2}{x}-(1-x)\ln_{2}{(1-x)}$ and $C$ denotes
the so-called Wooters concurrence 
\begin{equation}
C=\max{\{0,\sqrt{\lambda_{1}}-\sqrt{\lambda_{2}}-\sqrt{\lambda_{3}}-\sqrt{\lambda_{4}}\}}
\end{equation}
with $\lambda_{j}$'s indicating the eigenvalues of the matrix $\eta$
\begin{equation}
\eta=\sqrt{\rho_{1,1}}(\sigma_{y}^{A}\otimes\sigma_{y}^{B}){\rho_{1,1}}^{\ast}(\sigma_{y}^{A}\otimes\sigma_{y}^{B})\sqrt{\rho_{1,1}}\label{eq:autova4concurr}
\end{equation}
arranged in decreasing order. In the above expression ${\rho_{1,1}}^{\ast}$
denotes the complex conjugation of $\rho_{1,1}$, and $\sigma_{y}^{A(B)}$
is the well-known Pauli matrix acting on the qubit state controlled
by Alice(Bob).

The \emph{entanglement of particles} given in Eq.~\eqref{eq:entapart}
does not violate the superselection rule of the local particle number.
Indeed, local operations can be performed onto the collapsed state
$\rho_{1,1}$ without any restrictions in order to transfer its entanglement
$\mathcal{E}$ to the standard quantum registers of each subsystem.

Only recently, the estimation of quantum correlations (QCs) other
than entanglement has begun to be explored in systems of identical
particles by means of different approaches relying on the measurement-induced
disturbances~\cite{Majtey2013} or the use of witness operators~\cite{Iemini2013}.
By adopting the basic ideas of the approach by Wiseman and Vaccaro,
here we introduce a \emph{quantum} \emph{discord of particles} $D_{P}$.
Similarly to $E_{P}$, it represents the maximum amount of discord
that can be extracted from the quantum registers of distinguishable
qubits of Alice and Bob with no violation of the superselection rule
of the local particle number. Thus, $D_{P}$ can be expressed as:
\begin{equation}
D_{P}=P_{1,1}\mathcal{D}(\rho_{1,1}),\label{eq:discopart}
\end{equation}
where $\mathcal{D}$ quantifies the quantum correlations in the form
of discord between the Alice and Bob standard quantum registers in
the state $\rho_{1,1}$. It is given by\cite{Henderson2001,Ollivier2001}:
\begin{equation}
\mathcal{D}(\rho_{1,1})=\mathcal{T}(\rho_{1,1})-\mathcal{J}(\rho_{1,1}),
\end{equation}
that is the difference between the mutual information $\mathcal{T}$
and the classical correlation $\mathcal{J}$. Specifically, the former
can be expressed as: 
\begin{equation}
\mathcal{T}(\rho_{1,1})=S(\rho_{1,1}^{A})+S(\rho_{1,1}^{B})-S(\rho_{1,1})
\end{equation}
where $\rho_{1,1}^{A(B)}$ is the partial trace of the bipartite system
$\rho_{1,1}$ and $S(\rho_{1,1})=-\textrm{Tr}\left[\rho_{1,1}\log_{2}{(\rho_{1,1}})\right]$
is the von Neumann entropy. On the other hand, the amount of classical
correlation $\mathcal{J}$ is given by 
\begin{equation}
\mathcal{J}=\max_{\{\Pi_{k}^{B}\}}\left[S(\rho_{1,1}^{A})-S(\rho_{1,1}^{A}|\{\Pi_{k}^{B}\})\right],
\end{equation}
where $\{\Pi_{k}^{B}\}$ are projective measurements on subsystem
$B$ and $S(\rho^{A}|\{\Pi_{k}^{B}\})=\sum_{k}p_{k}\rho_{k}^{A}$,
where $\rho_{k}^{A}=\text{Tr}_{B}[\Pi_{k}^{B}\rho\Pi_{k}^{B}]/\text{Tr}[\Pi_{k}^{B}\rho\Pi_{k}^{B}]$
is the density operator describing $A$ conditioned by the measurement
outcome $k$ on $B$. 

It is worth noting that, from the definitions themselves, both $E_{P}$
and $D_{P}$ depend upon the partition of the system, that is upon
which modes Alice and Bob control.

Considering the definition of Eq. \eqref{eq:discopart}, together
with the invariance of $P_{1,1}$ and $\mathcal{D}(\rho_{1,1})$ under
local transformations (which conserve the local particle number),
it is easy to prove that the \emph{discord of particles} possesses
all the required properties for a quantum discord quantifier\cite{Henderson2001,Modi2012},
as shown in Appendix~\ref{sec:Appendix_QD}. This means that, like
the conventional \emph{quantum discord}, our quantifier should be
able to capture some features of quantum correlations among particles
that are missed by the \emph{entanglement of particles}.

\section{Physical models\label{sec:Physical-models}}

The Hubbard model represents a valid mean to describe a number of
different systems in solid-state physics ranging from ultracold atoms
trapped in optical lattices to high temperature superconductivity~\cite{essler,Amico2008}.
Since its physical relevance, it is of interest to examine the dynamics
of quantum correlations in simplified Hubbard models of interacting
bosons and fermions subject to a classical external noise. Specifically,
we focus on Bose- and Fermi-Hubbard models affected by environmental
RTN, pink $1/f$, and brown $1/f^{2}$ noises.

\subsection{Two-boson Bose-Hubbard model\label{sub:BoseHubb}}

\begin{figure}
\includegraphics[scale=0.45]{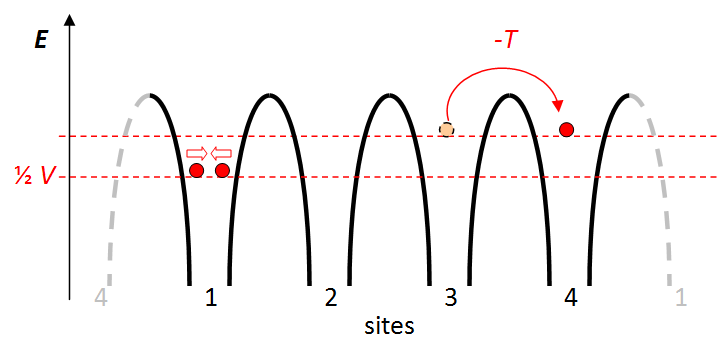}

\caption{Bosonic Hubbard chain: $V$ is the interaction strength between particles
sharing the same site, $-T$ is the energy gain for hopping (in figure
$T>0$ and $V<0$).}
\label{fig:BOSHubbChain}

\end{figure}

Here, we consider two spinless bosons interacting among each other
and hopping among four sites of a lattice. This system is the bosonic
version of the Hubbard plaquette, which is used in solid state as
a building block for highly-correlated many-body systems\citep{Kim1999,*Altman2002,*Tsai2008,*Gull2008}.
In agreement with previous works~\cite{Buscemi2013,Benedetti2013,Bordone2012,Benedetti2012},
in order to mimic the effect of the noise on the quantum system the
hopping amplitudes among the sites are assumed to follow a time-dependent
stochastic behavior. Such an assumption yields the Hamiltonian~\cite{Dowling2006,Lahini2012}
\begin{align}
\mathcal{H}_{BH}(t)=-T\sum_{i=1}^{4}q_{i}(t)(b_{i}^{\dag}b_{i+1}+b_{i+1}^{\dag}b_{i})\nonumber \\
+\frac{V}{2}\sum_{i=1}^{4}\hat{n}_{i}(\hat{n}_{i}-1),\label{eq:BosHubbHam}
\end{align}
where $b_{i}^{\dag},b_{i},$ are the bosonic creation and annihilation
operators for a particle at site $i$ (satisfying the commutation
rules $[b_{i}^{\dag},b_{j}]=\delta_{ij}$) in the Fock space, $\hat{n}_{i}=b_{i}^{\dag}b_{i}$
is the corresponding number operator, and we have imposed periodic
boundary conditions (PBC), namely $i+1=1$ for $i=4$. $V$ denotes
the on-site interaction energy, $T$ indicates the tunneling amplitude
between the neighbor sites in absence of noise, and $q_{i}(t)$ is
a time-dependent random parameter related to the kind of noise.

For the case of random telegraph noise (RTN), $q_{i}(t)=\eta_{i}(t)$
where $\eta_{i}(t)$ describes a single fluctuator randomly flipping
between the values $-1$ and $1$ at rate $\gamma_{0}$. In our approach,
we assume that the sources of noise affecting the tunneling among
neighbor sites are independent of each other, that is $q_{0}(t)\neq q_{1}(t)\neq q_{2}(t)\neq q_{3}(t)$.
On the other hand, for the pink and Brownian noise, $q_{i}(t)=\frac{1}{N_{f}}\sum_{j=1}^{N_{f}}\eta_{ij}(t)$
is given by an averaged linear superposition of $N_{f}$ bistable
fluctuators $\eta_{ij}(t)$, each one with a proper switching rate
$\gamma_{j}$ taken from the range $[\gamma_{\textrm{inf}},\gamma_{\textrm{sup}}]$
according to the distribution 
\begin{equation}
p(\gamma)=\left\{ \begin{array}{cc}
\frac{1}{\gamma}\frac{1}{\ln{(\gamma_{\textrm{sup}}/\gamma_{\textrm{inf}})}} & \quad\textrm{for}\,1/f\,\textrm{noise}\\
\\
\frac{1}{\gamma^{2}}\frac{1}{1/\gamma_{\textrm{inf}}-1/\gamma_{\textrm{sup}}} & \quad\quad\textrm{for}\,1/f^{2}\,\textrm{noise.}
\end{array}\right.
\end{equation}

As shown elsewhere~\cite{Benedetti2013}, the greater is $N_{f}$,
the closer is the power spectrum of the stochastic process described
by $\sum_{j=1}^{N_{f}}\eta_{ij}(t)$ to the $1/f^{\alpha}$-like behavior
(where $\alpha$=1,2), in a frequency interval $[f_{1},f_{2}]$ with
$\gamma_{\textrm{inf}}\ll f_{1},f_{2}\ll\gamma_{\textrm{sup}}$. Here,
we average the effect of the $N_{f}$ bistable fluctuators in order
to keep the hopping amplitude $T_{i}(t)=T\cdot q_{i}(t)$ in the interval
$[-|T|,|T|]$, since - as we will see briefly - the absolute value
of $T$ determines the characteristic speed of the evolution of the
system.

For both kinds of noise, given the random nature of the Hamiltonian
of Eq.~\eqref{eq:BosHubbHam}, the time evolution of the two-boson
system is obtained by averaging the stochastic dynamics of the quantum
state over different noise configurations. Specifically, we adopt
a numerical approach able to generate, for each noise parameter $q_{i}(t)$,
a given number $M$ of histories (i.e., different temporal sequences
of RTN or $1/f^{\alpha}$ noise signals) which are inserted in $\mathcal{H}_{BH}(t)$
to evaluate $M$ unitary time evolutions of the system. Once estimated
these, the two-boson density matrix $\rho(t)$ at time $t$ can be
evaluated as: 
\begin{equation}
\rho(t)=\frac{1}{M}\sum_{k=1}^{M}U_{k}(t)\rho(0)U_{k}^{\dag}(t),
\end{equation}
where $U_{k}(t)$ is the time-evolution operator corresponding to
the $k$-th history, and the initial state of the system $\rho(0)$
is 
\begin{align}
\rho_{B}(0) & =\left|\Psi_{B}\right\rangle \left\langle \Psi_{B}\right|,
\end{align}
where
\begin{equation}
\left|\Psi_{B}\right\rangle =\frac{1}{\sqrt{2}}\bigg(b_{3}^{\dag}b_{1}^{\dag}+b_{4}^{\dag}b_{2}^{\dag}\bigg)|0\rangle\label{eq:Psi_B}
\end{equation}
and $|0\rangle$ indicating the vacuum state containing zero particles
in each mode. Then, for the bipartition of the system $A=\{1,2\}$
and $B=\{3,4\}$, $\left|\Psi_{B}\right\rangle $ is a two-boson maximally
entangled state. Here, we address the decohering effects suffered
by the two-boson entanglement and discord between the two-site bipartitions
$A$ and $B$.

\subsection{Two-fermion Fermi-Hubbard model\label{sub:Fermi-Hubb}}

\begin{figure}
\includegraphics[scale=0.45]{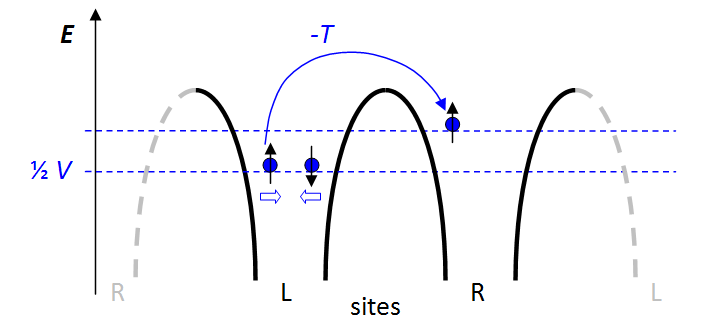}

\caption{Fermionic Hubbard dimer: $V$ is the interaction strength between
electrons sharing the same site, $-T$ is the energy gain for hopping
(in figure $T>0$ and $V<0$).}
\label{fig:FERHubbChain}
\end{figure}

Here, we illustrate a Hubbard dimer of two electrons with spin degrees
of freedom $\sigma=\uparrow,\downarrow$ hopping between two spatial
sites $L$ and $R$ in the presence of noise. This simple model allows
for the description of a large number of systems\cite{Kajala2011,giamarchi2004quantum}
– among which there is the hydrogen molecule – since it is strictly
connected to the antiferromagnetic Heisenberg model\cite{Zanardi2002}
in the high interaction limit.

The two-fermion dynamics is ruled by the stochastic Hamiltonian: 
\begin{eqnarray}
\mathcal{H}_{HD}(t)=-T\sum_{\sigma=\uparrow,\downarrow}q_{\sigma}(t)(c_{L\sigma}^{\dag}c_{R\sigma}+c_{R\sigma}^{\dag}c_{L\sigma})\nonumber \\
+\frac{V}{2}\sum_{i=L,R}\hat{n}_{i\uparrow}\hat{n}_{i\downarrow},\label{eq:Fermi_Hubb_Hamilt}
\end{eqnarray}
where $c_{i\sigma}^{\dag},c_{i\sigma}$ are the creation and annihilation
operators of a fermion at site $i$ with spin $\sigma$ satisfying
the anticommutation rules $\{c_{i\sigma}^{\dag},c_{i^{\prime}\sigma^{\prime}}^{\dag}\}=\delta_{i,i^{\prime}}\delta_{\sigma,\sigma^{\prime}}$.
$T$ still denotes the hopping amplitude between the spatial sites
while the $V$ term mimics a sort of Coulomb interaction between electrons
on the same site. The RTN, pink, and brown noises are reproduced by
means of the random term $q_{\sigma}(t)$ which is assumed to be dependent
upon spin. Unlike the Bose-Hubbard model, the electron hopping among
spatial sites is only affected by two different sources of noise,
and spin-flip transitions are prohibited.

The numerical procedure described in the Sec.~\ref{sub:BoseHubb}
is again adopted to evaluate the time-evolution of the two-fermion
state 
\begin{align}
\rho_{F}(0) & =\left|\Psi_{F}\right\rangle \left\langle \Psi_{F}\right|,
\end{align}
where
\begin{equation}
\left|\Psi_{F}\right\rangle =\frac{1}{2}\bigg(c_{L\uparrow}^{\dag}c_{L\downarrow}^{\dag}+c_{R\uparrow}^{\dag}c_{R\downarrow}^{\dag}\bigg)|0\rangle.\label{eq:Psi_F}
\end{equation}
This is a maximally-entangled state when the bipartition of the system
$A=\{L\uparrow,R\uparrow\}$ and $B=\{L\downarrow,R\downarrow\}$
is considered. Indeed, our aim is to evaluate how the noise affects
the quantum correlations between the spins up and down of the two
electrons.

\subsection{General remarks on the Hubbard Hamiltonian\label{sub:General-remarks-HB}}

With a simple factorization, the Bose-Hubbard Hamiltonian of Eq. \eqref{eq:BosHubbHam}
can be rewritten as:

\begin{align}
\mathcal{H}_{BH}(t)= & T\left[-\sum_{i=1}^{4}q_{i}(t)(b_{i}^{\dag}b_{i+1}+b_{i+1}^{\dag}b_{i})\right.\label{eq:HamBH_comp}\\
 & \left.+\frac{v}{2}\sum_{i=0}^{3}\hat{n}_{i}(\hat{n}_{i}-1)\right],\nonumber 
\end{align}
where $v=V/T$ is the relative strength of on-site interaction with
respect to the kinetic term. As we can see, the physical meaning of
$T$ can be reduced to a time-scale factor, i.e. the higher is $T$,
the faster is the dynamics of the system (so the evolution of the
correlations has to be evaluated against the adimensional time $\tau=\left|T\right|\cdot t$).
Therefore, the only parameter which is able to change significantly
the dynamics of the system is $v$, and analogous conclusions can
be drawn for the Fermi-Hubbard Hamiltonian. One could expect that
the sign of $V$, that distinguishes between attractive ($V<0$) and
repulsive ($V>0$) interactions among particles, should alter dramatically
the dynamics of the system. This, however, seems in contrast with
the literature, where it is shown that the evolution of the system
is invariant with respect to the sign of $V$ \cite{Lahini2012}.
Indeed, after a closer look, it turns out that the dynamics of quantum
correlations (QC) is independent from the sign of $V$ only when the
system is in a state of the Fock space with exact occupation numbers
for each site (the case studied in Ref. \cite{Lahini2012}), while
for linear combinations of states the dynamics can be affected by
sign of interactions in a non negligible way, as we will show briefly.

This point will be rigorously detailed in a forthcoming work.

\section{Numerical simulations: Bose-Hubbard model\label{sec:Numerical-simulations:-Bose-Hubb}}

In this Section we report the results of our numerical simulations.
Our algorithm generates the time-dependent Hamiltonian $\mathcal{H}(t)$
at time steps of $\delta t$, updating the values of the hopping amplitudes
according to the random noise (if present), then it calculates the
evolution $\rho(t)=U(t)\rho(0)U^{\dagger}(t)$ of the initial state
$\rho(0)$ through the time-ordered evolution operator 
\begin{equation}
U(t)=\mathcal{T}\left\{ \exp\left[-i\int_{0}^{t}\mathcal{H}(\tau)d\tau\right]\right\} 
\end{equation}
 at desired time $t=n\cdot\delta t$. The operator $U(t)$ is discretized
in time and written as\cite{curotto2009stochastic} 
\begin{equation}
\mathcal{T}\left\{ \exp\left[-i\sum_{j=0}^{n}\mathcal{H}(j\cdot\delta t)\delta t\right]\right\} \simeq\prod_{j=0}^{n}\exp\left[-i\mathcal{H}(j\cdot\delta t)\delta t\right]
\end{equation}
resorting to the Trotter factorization in the limit of $\delta t\rightarrow0$,
and each term $\exp\left[-i\mathcal{H}(j\cdot\delta t)\delta t\right]$
is obtained through the diagonalization of $\mathcal{H}(j\cdot\delta t)$.
The evolution is calculated in the joint Hilbert space of the two
particles (symmetrization or anti-symmetrization of the wavefunction
can be performed equally at the beginning of the evolution or at each
time-step). The values of $\rho(t)$ are then averaged over a number
of histories which is large enough to ensure convergence, and the
average density matrix $\left\langle \rho(t)\right\rangle $ is used
to evaluate entanglement $E_{P}$ and discord $D_{P}$ of particles
at time $t$, as explained in Sect.~\ref{sec:Bipartite-entanglement-and-discord}.
The numerical optimization of quantum discord $\mathcal{D}(\rho_{1,1})$
is performed with the \emph{exhaustive enumeration} algorithm\emph{
}(aka brute-force search), in order to reach a global solution.

\subsection{Noiseless system\label{sub:BOS-Noiseless-system}}

\begin{figure*}[!tp]
\includegraphics[scale=0.45]{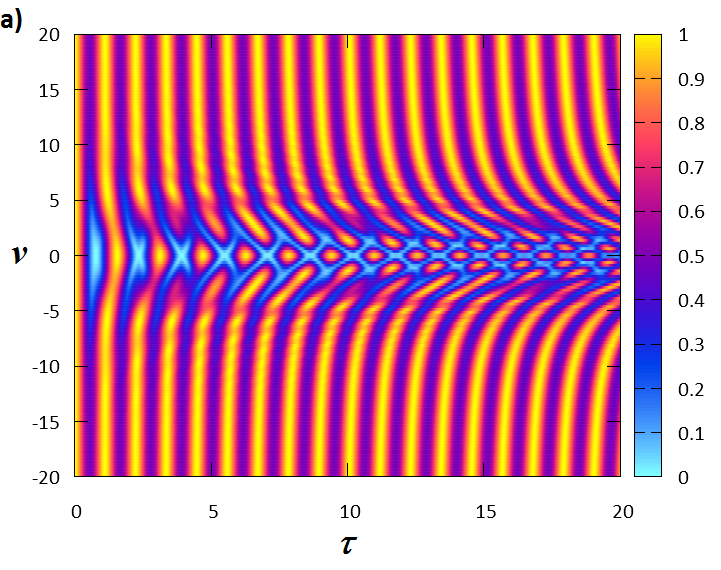}\hfill \includegraphics[scale=0.45]{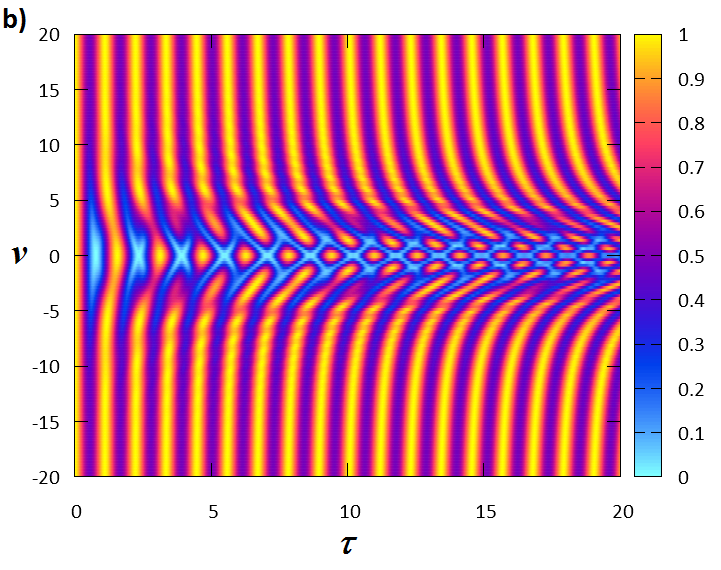}

\caption{Entanglement of particles and Discord of particles for the initial
state $\left|\Psi_{B}\right\rangle $. \textbf{a)} Entanglement $E_{P}$
\textbf{b)} Discord $D_{P}$.}
\label{fig:ED_bos_noiseless_Psi}

\begin{minipage}[t]{.49\textwidth}
\begin{centering}

\includegraphics[scale=0.4]{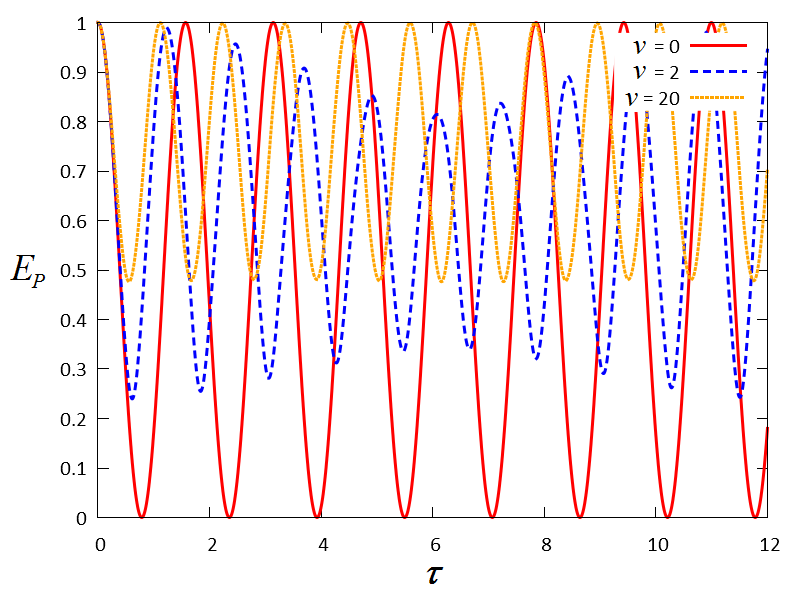}

\caption{Entanglement of particles for the initial state $\left|\Psi_{B}\right\rangle $
at different relative strength $v$ of interactions.}
\label{fig:E_bos_noiseless_Psi_oscill}

\end{centering}
\end{minipage}
\hfill
\begin{minipage}[t]{.49\textwidth}
\begin{centering}

\includegraphics[scale=0.4]{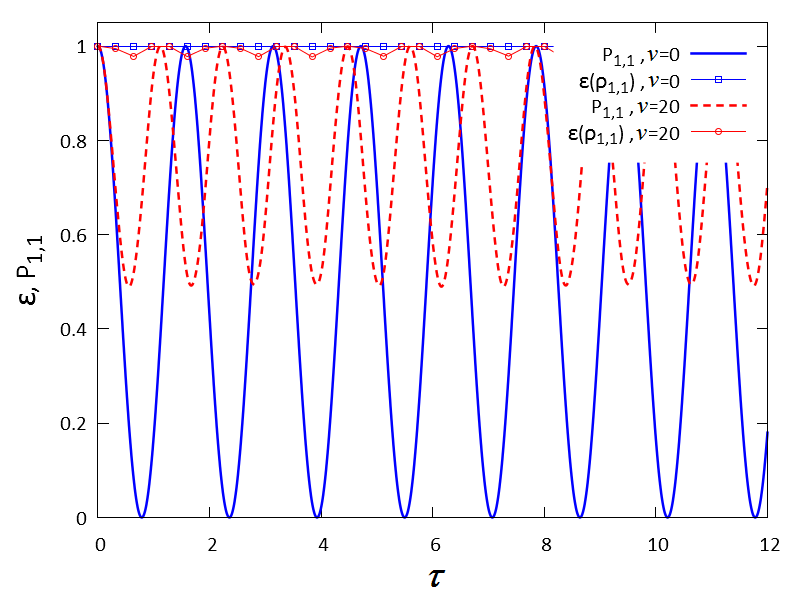}

\caption{Entanglement of formation $\mathcal{E}(\rho_{1,1})$ and probability
$P_{1,1}$ of finding the particles in different partitions for the
initial state $\left|\Psi_{B}\right\rangle $ at different relative
strength $v$ of interactions.}
\label{fig:E_bos_noiseless_P11_E11}

\end{centering}
\end{minipage}

\begin{minipage}[t]{.49\textwidth}
\begin{centering}

\includegraphics[scale=0.45]{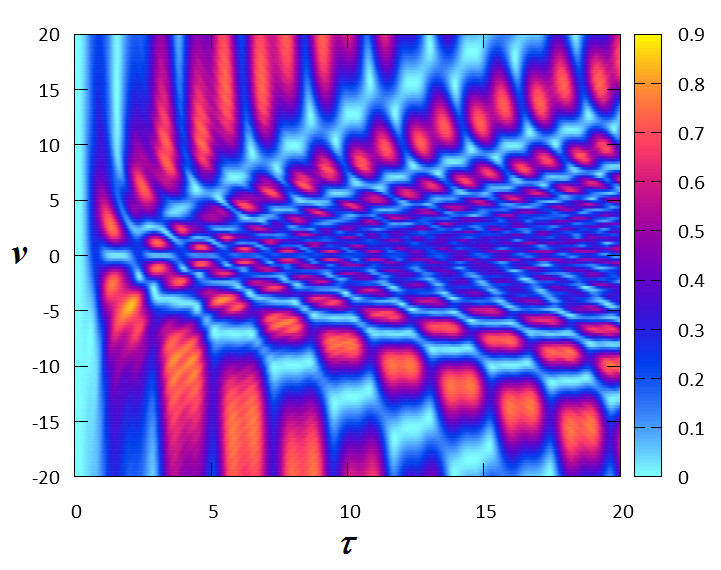}

\caption{Entanglement of particles $E_{P}$ for the initial state $\left|\Xi_{B}\right\rangle $
(see text for definition). The results for Discord of particles $D_{P}$
are not reported since they are identical.}
\label{fig:E_bos_noiseless_Xi-1}

\end{centering}
\end{minipage}
\hfill
\begin{minipage}[t]{.49\textwidth}
\begin{centering}

\includegraphics[scale=0.4]{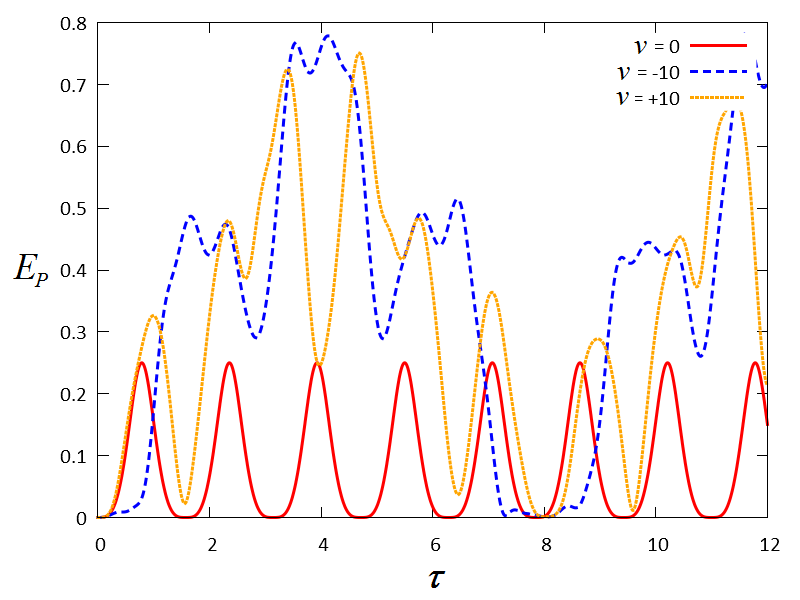}

\caption{Entanglement of particles for the initial state $\left|\Xi_{B}\right\rangle $
(see text for definition) at different relative strength $v$ of interactions.}
\label{fig:E_bos_noiseless_Xi_v}

\end{centering}
\end{minipage}
\end{figure*}

The numerical simulations for the noiseless system give the following
results. We start with the maximally entangled state $\left|\Psi_{B}\right\rangle $
of Eq.\eqref{eq:Psi_B}, and the evolution of the correlations are
practically identical for Entanglement $E_{P}$ (Fig. \ref{fig:ED_bos_noiseless_Psi}
a) and Quantum Discord $D_{P}$ (Fig. \ref{fig:ED_bos_noiseless_Psi}
b), thus showing that our quantifier is qualitatively able to capture
the quantum correlations of the noiseless system.

Both entanglement and discord show a periodic behavior in time (sometimes
with beatings - see Fig. \ref{fig:E_bos_noiseless_Psi_oscill}). This
is perfectly reasonable since the Hilbert space of the system is finite
and the noiseless Hamiltonian is constant in time, therefore the dynamics
can repeat itself after a certain period of time (provided that the
involved eigenvalues have rational quotients). For longer chains,
where the effect of the PBC shows up at later times, the period of
the dynamics would be obviously larger.

For what concerns the effects of the potential energy $V$, we see
that at larger values of $v$ the oscillations of the quantum correlations
are smaller in amplitude and characterized by a higher average value
(see Fig. \ref{fig:E_bos_noiseless_Psi_oscill}), thus showing a sort
of ``protective'' effect of the interactions with respect to both
entanglement and discord. As we can see from both Fig. \ref{fig:ED_bos_noiseless_Psi}
a) and b), QC are independent from the sign of $V$, as already observed
for both bosonic\cite{Lahini2012} and fermionic systems\cite{schneider2010}.
Interestingly, however, this effects is not universal but it depends
on the chosen initial state. Indeed, for a different initial state,
such as $\left|\Xi_{B}\right\rangle =\frac{1}{\sqrt{2}}(b_{4}^{\dag}b_{1}^{\dag}+b_{4}^{\dag}b_{2}^{\dag})|0\rangle$,
the correlations show an asymmetry%
\footnote{The same asymmetry is observed for the diagonal elements of $\rho'_{B}(t)=|\Xi_{B}{\rangle}{\langle}\Xi_{B}|$,
while it is absent in $\rho_{B}(t)=|\Psi_{B}{\rangle}{\langle}\Psi_{B}|$,
but we do not report this result here for reasons of space: however,
we mention this phenomenon since it is reproduced also for larger
chains (e.g. with 30 sites) and for couples of particles initialized
far from the edges of the chain, thus showing that this effect is
genuine and not due to PBC.%
} with respect to $V=0$ (see Fig. \ref{fig:E_bos_noiseless_Xi-1}).

In this second case, even if the initial entanglement is zero (since
the state can be factorized), correlations do appear in any case,
and not only because of the effect of interactions (which can entangle
the particles), but also because of periodic boundary conditions:
indeed, even when $v=0$ (see Fig. \ref{fig:E_bos_noiseless_Xi_v}),
the interference effects can build up nonclassical correlations between
particles, as already observed in the literature \cite{Franson2013,SamuelssonButtiker_ABEffect,Benedetti2012_QWferbos}.
Also, no periodicity is apparent for the observed cases, except for
$V=0$, but we observe again a larger average value of QC for higher
values of $v$.

\begin{figure*}[!t]
\includegraphics[scale=0.4]{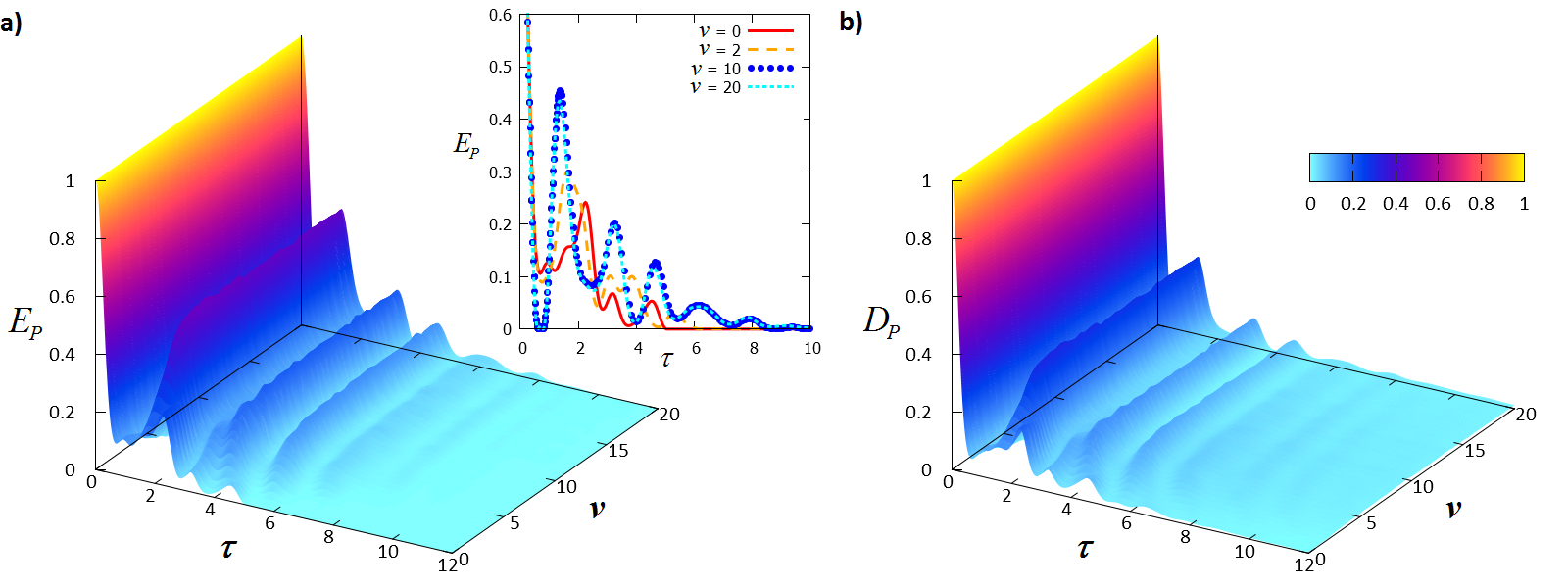} 

\caption{Entanglement $E_{P}$ and Discord of particles $D_{P}$ with a single
RTN fluctuator in the strong coupling regime ($\gamma_{0}\tau_{s}=0.1$).
\textbf{Inset:} $E_{P}$ for different values of the interaction strength
$v$.}
\label{fig:ED_Bos_RTN_nmk}

\includegraphics[scale=0.4]{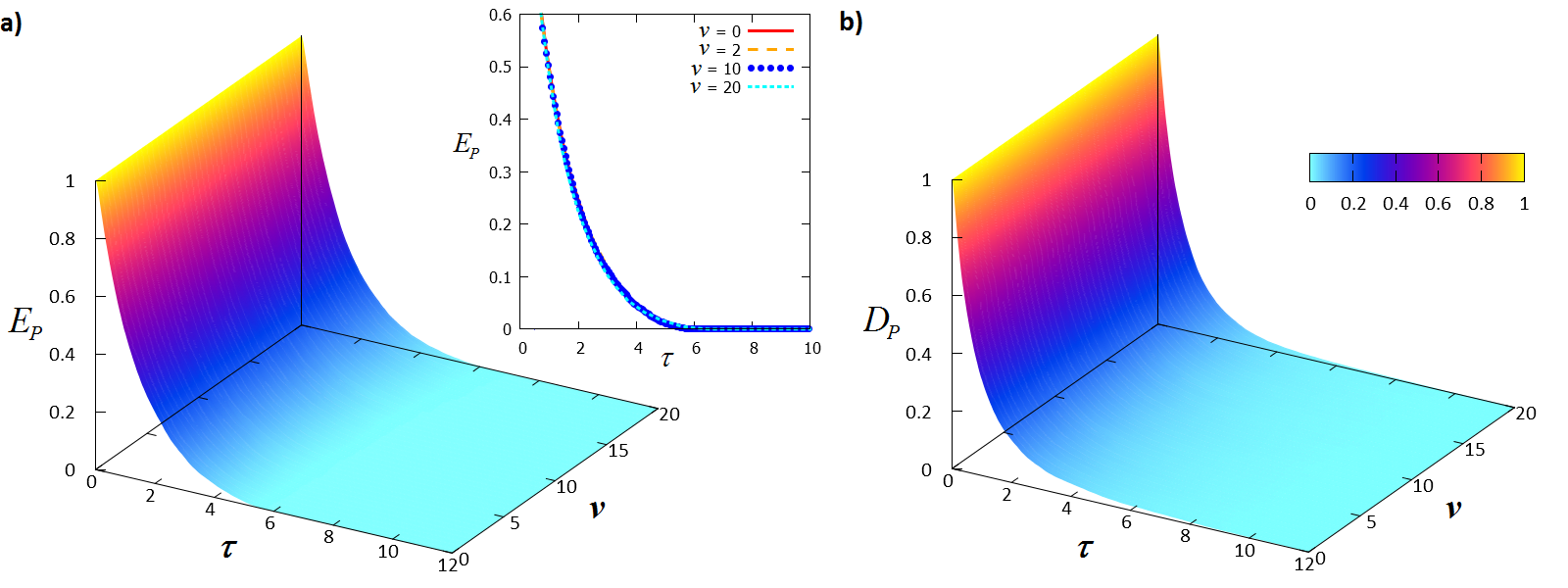}

\caption{Entanglement $E_{P}$ and Discord of particles $D_{P}$ with a single
RTN fluctuator in the weak coupling regime ($\gamma_{0}\tau_{s}=10$).
\textbf{Inset:} $E_{P}$ for different values of the interaction strength
$v$.}
\label{fig:ED_Bos_RTN_mk}
\end{figure*}

However, since we are mainly interested in the evolution of a maximally
entangled state, from now on we will focus only on the effects of
the noise on the correlations of the state $\left|\Psi_{B}\right\rangle $.
As we will see, the QC for $\left|\Psi_{B}\right\rangle $ are always
symmetrical with respect to $v$ (but this result should be considered
with care since - as we just saw - this is not a general property
for the system).

Looking back at Eq. \eqref{eq:entapart}, we recall that entanglement
of particles $E_{P}$ is the product of two contributions: the first
is the probability of finding one particle in each partition, namely
$P_{1,1}$, and the second is the entanglement of modes for the standard
quantum registers, namely $\mathcal{E}(\rho_{1,1})$. The role of
these contributions is compared in Fig. \ref{fig:E_bos_noiseless_P11_E11}.
As we can see, the entanglement of modes $\mathcal{E}$ has very low
dependence on the relative strength $v$ of the interactions. On the
contrary, the probability $P_{1,1}$ depends strongly on $v$: for
non-interacting particles it oscillates between 0 and 1, while for
strongly interacting particles ($v=20$) it is bounded between 0.5
and 1, thus producing the strong increase in the average value of
$E_{P}$ that we observe in Fig. \ref{fig:E_bos_noiseless_Psi_oscill}
for high values of $v$. This means that the role of strong interactions
is that of keeping preferentially both particles in their original
partition, or at least discouraging two particles from occupying the
same partition of the system. From a physical point of view, this
can be connected to the band-structure of the Hubbard Hamiltonian\cite{Lahini2012}.
Indeed, as observed also in Mott-Hubbard transitions\cite{altland2010condensed},
the increase in $v$ separates the bandstructure into two subbands:
a main one, for states in which the particles are on first-neighbor
or second-neighbor sites (e.g. $\left|1,2\right\rangle $ and $\left|1,3\right\rangle $),
and a miniband, for states where the particles share the same site
(e.g. $\left|1,1\right\rangle $). Given our initial state $\left|\Psi_{B}\right\rangle $,
the separation in energy of the subbands lowers the probability of
certain transitions for the particles, namely those in which they
occupy the same site (and therefore they are in the same partition),
thus increasing the relative probability of occupying different partitions.
Obviously, the same discussion holds for the discord of particles
$D_{P}$.

\subsection{Single RTN fluctuator\label{sub:BOS-Single-RTN-fluctuator}}

When the hopping amplitudes of the particles are perturbed each one
by a single RTN fluctuator with switching rate $\gamma_{0}$, two
regimes of behavior can be identified for the system: the strong coupling
regime $\gamma_{0}\tau_{s}<1$, in which the switching dynamics of
the fluctuators are slower than the characteristic time $\tau_{s}=|T|^{-1}$of
the system (thus allowing a back-action of the environment over the
system), and the weak coupling regime $\gamma_{0}\tau_{s}>1$, where
the previous condition is reversed, and the effect of the environment
is perceived by the system as an average over many noise cycles.

All the simulations for the initial state $\left|\Psi_{B}\right\rangle $
show that the results are independent from the sign of $V$, therefore
we report the values of QC only for $v>0$. For what concerns the
strong coupling regime, as we see from Fig. \ref{fig:ED_Bos_RTN_nmk},
the qualitative behavior for entanglement and discord is the same,
and the system shows a sort of memory effect, with sudden deaths and
revivals of correlations before reaching a complete decoherence condition.
However, discord seems to be more sensitive to the noisy environment,
since it assumes typically lower values with respect to entanglement
after the switching on of the noise (except for long times, when a
crossing of the two curves can occur before they both go to zero).
This result is not surprising \cite{Luo2008a}, since it is know that
the two quantifiers do capture different kinds of quantum correlations,
and therefore a direct quantitative comparison between them is not
possible. Again, higher values of $v$ have a protective effects on
correlations, that is they disappear at longer times. However, for
$v\geq10$, a sort of limit behavior is reached, and higher values
of $v$ do not produce any modification in QC (see inset of Fig. \ref{fig:ED_Bos_RTN_nmk}).

A different behavior of QC is observed in the weak coupling regime
(see Fig. \ref{fig:ED_Bos_RTN_mk}), where discord is still lower
than entanglement (except for long times, where a crossing can again
occur), but the variation of $v$ does not produce any change in the
time evolution of discord or entanglement (see inset in Fig. \ref{fig:ED_Bos_RTN_mk}).
Under this coupling regime, the system does not show memory effects,
and the decay of QC is monotone, but not necessarily faster than that
of the strong coupling regime. These behaviors are very similar to
those observed in a couple of qubits subject to RTN in the Markovian
and non-Markovian regime \cite{Benedetti2012}, except for the decay
times of correlations, which are apparently faster for qubits in the
Markovian regime.

\subsection{Colored noises\label{sub:BOS-Colored-noises}}

\begin{figure*}[p]
\includegraphics[scale=0.4]{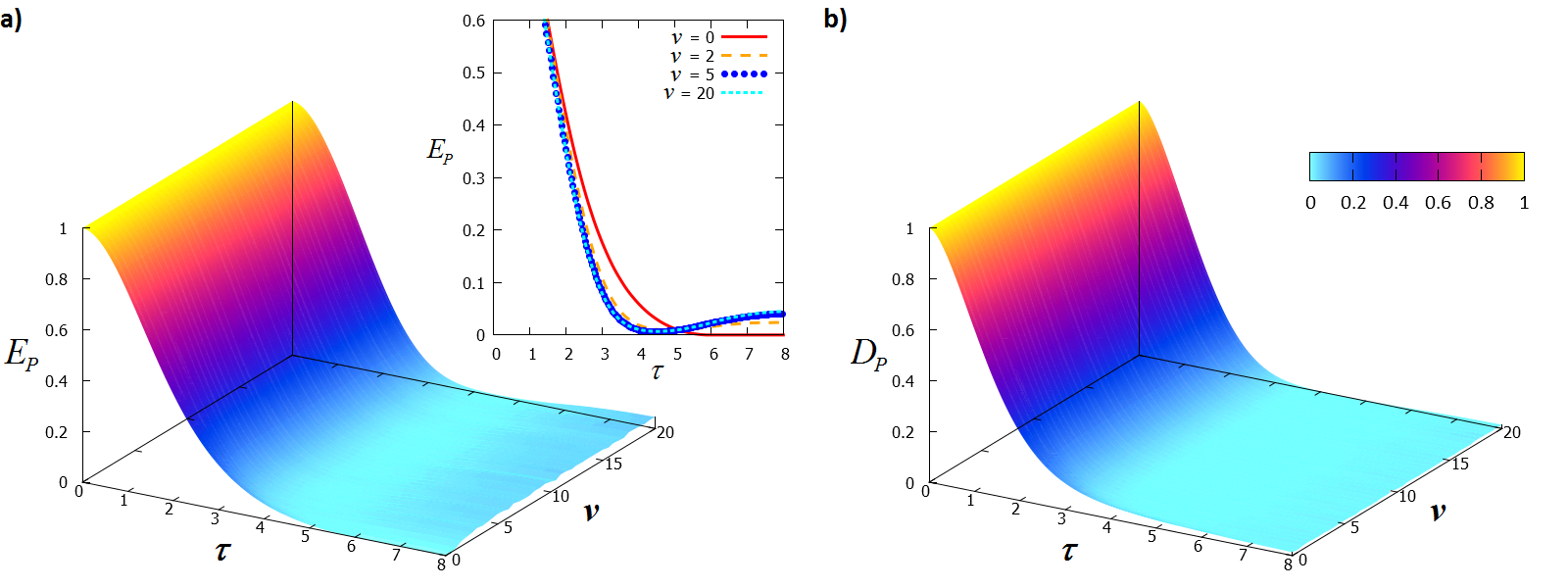} 

\caption{Entanglement $E_{P}$ and Discord of particles $D_{P}$ for $1/f$
(pink) noise, with $N_{f}=20$ fluctuators. \textbf{Inset}: Evolution
of $E_{P}$ for different strengths $v$ of the interactions.}
\label{fig:ED_BOS_PinkNoise}

\includegraphics[scale=0.4]{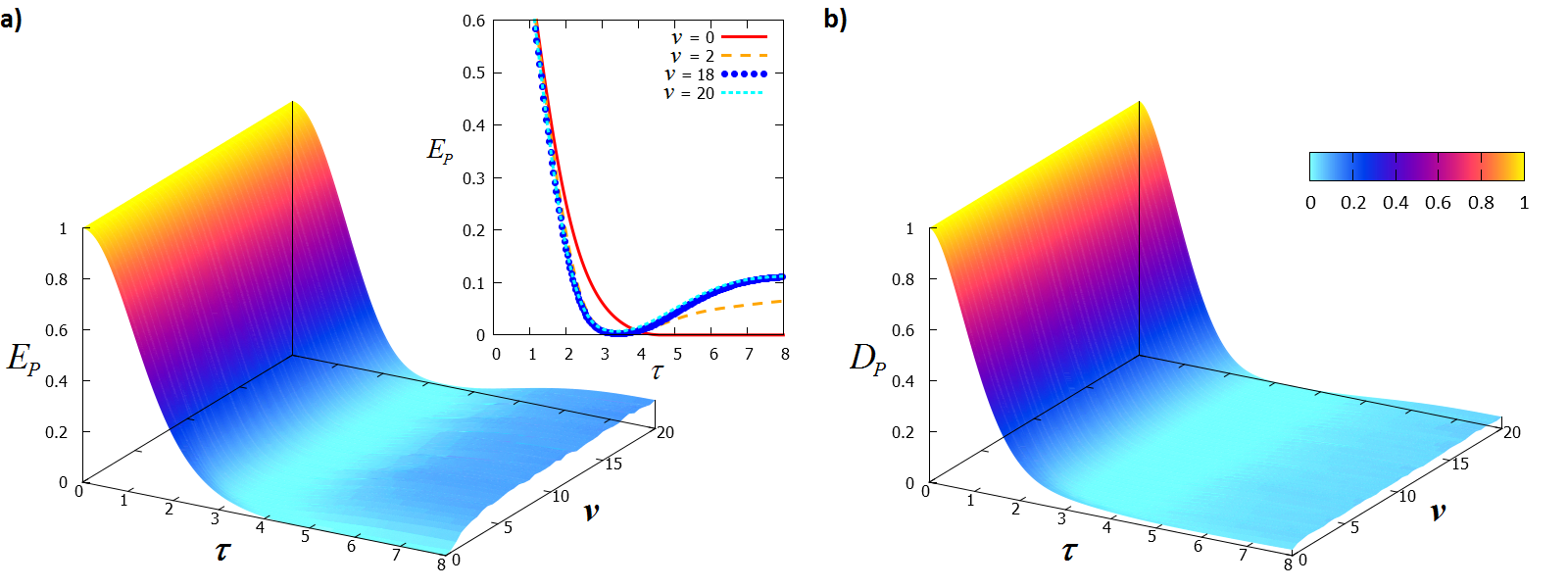} 

\caption{Entanglement $E_{P}$ and Discord of particles $D_{P}$ for $1/f^{2}$
(brown) noise, with $N_{f}=20$ fluctuators. \textbf{Inset}: Evolution
of $E_{P}$ for different strengths $v$ of the interactions.}
\label{fig:ED_BOS_BrownNoise}

\vspace{5mm}

\begin{minipage}[t]{.49\textwidth}
\begin{centering}

\includegraphics[scale=0.4]{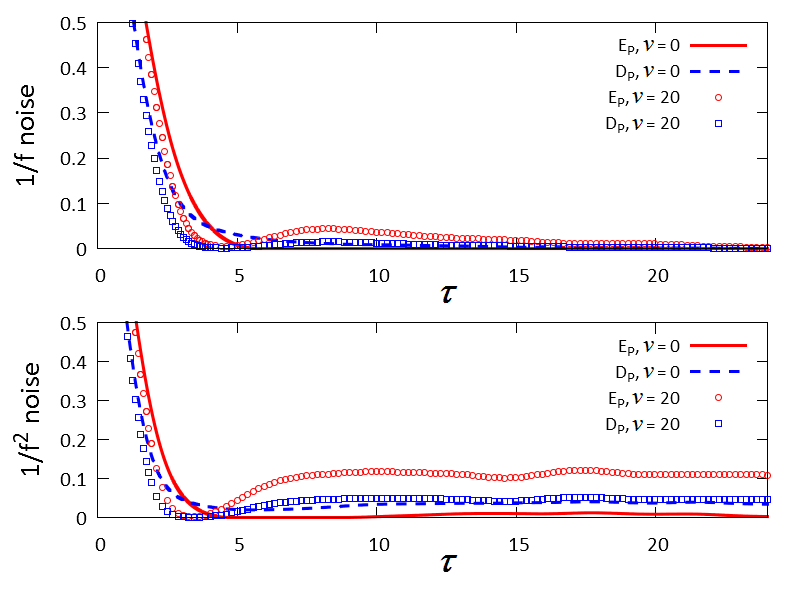} 

\caption{Entanglement $E_{P}$ (red) and Discord $D_{P}$ (blue) for $1/f$
(top panel) and $1/f^{2}$ (bottom panel) noise, with $N_{f}=20$
fluctuators, at different values of the interaction strength $v$
and for long time scales.}
\label{fig:ED_BOS_PinkBrown_with_v}

\end{centering}
\end{minipage}
\hfill
\begin{minipage}[t]{.49\textwidth}
\begin{centering}

\includegraphics[scale=0.4]{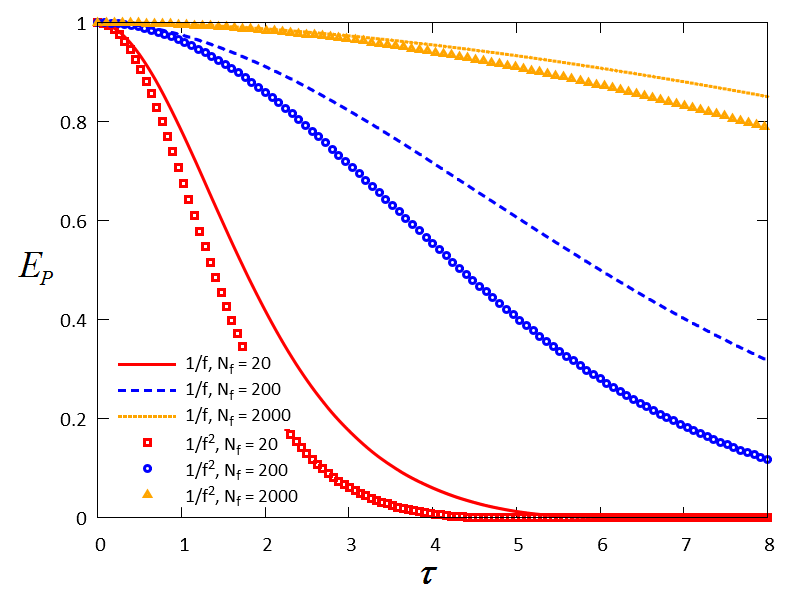}

\caption{Entanglement of particles for $1/f$ (lines) and $1/f^{2}$ (dots)
noise at zero interaction strength, for different values in the number
of fluctuators $N_{f}$.}
\label{fig:ED_BOS_PinkBrown_with_Nf}

\end{centering}
\end{minipage}
\end{figure*}

If the hopping amplitudes are perturbed by a collection of $N_{f}$
bistable fluctuators (each of them mimicking a decohering channel),
whose switching frequencies $\gamma=f$ are distributed with the power
law $f^{-\alpha}$, then the system is subject to pink ($\alpha=1$)
or brown ($\alpha=2$) colored noise. To reproduce these kinds of
noise, we generated randomly the rates $\gamma$ of our fluctuators
in the interval of frequencies $\gamma\in|T|\cdot[1.25\cdot10^{-4},1.25\cdot10^{2}]$,
using the appropriate power law for each noise. Again, since the simulations
for the initial state $\left|\Psi_{B}\right\rangle $ show that the
results are independent from the sign of $V$, we report the evolution
of QC only for $v>0$. We notice an interesting dependence of both
entanglement and discord from the absolute values of the interaction
$v$, both for $1/f$ (Figs. \ref{fig:ED_BOS_PinkNoise} a and b)
and $1/f^{2}$ noises (Figs. \ref{fig:ED_BOS_BrownNoise} a and b).
In detail, while QC for $v=0$ show their first decay at long times,
with $v>0$ they decay faster but show a marked revival (see the insets
of Figs. \ref{fig:ED_BOS_PinkNoise} and \ref{fig:ED_BOS_BrownNoise}).
These phenomena are present in both pink and brown noise scenario,
but they are more marked for brown noise (Fig. \ref{fig:ED_BOS_PinkBrown_with_v}),
where QC at long times reach an almost constant value (see Fig. \ref{fig:ED_BOS_PinkBrown_with_v}).
This revival is stronger for higher values of $v$, when an asymptotic
behavior is reached – e.g., in the case of $E_{P}$ the saturation
is given by $v\gtrsim5$ for $1/f$ noise, $v\gtrsim18$ for $1/f$
noise (see insets of Figs. \ref{fig:ED_BOS_PinkNoise} and \ref{fig:ED_BOS_BrownNoise}).
Again, interactions show to have a protective effect on QC. However,
we also notice that quantum discord $D_{P}$, although it decays at
first generally faster than $E_{P}$ and has weaker revivals, at $v=0$
manages to overcome entanglement: in the pink noise scenario it happens
before they both go to zero, while in the brown noise scenario discord
never fully vanishes (see Fig. \ref{fig:ED_BOS_PinkBrown_with_v}),
while entanglement goes rapidly to zero (and shows a negligible revival).
We comment further on this behavior of entanglement and discord in
Appendix \ref{sec:Appendix_Pur}.

The decohering effects strongly depends upon the number $N_{f}$ of
fluctuators, and if the sum $q(t)$ of the fluctuators is not renormalized,
the decohering effect is stronger for a higher number $N_{f}$ of
open noise channels (also because this produces a larger average value
of $T(t)$, and therefore a faster dynamics for the system), as it
is shown in \cite{Benedetti2013}. However, since in our simulations
we renormalize the noise $q(t)$ to the number of fluctuators, the
effect we observe here is reversed: in Fig. \ref{fig:ED_BOS_PinkBrown_with_Nf}
we see that for higher values of $N_{f}$ a longer time is needed
to destroy the QC. This is reasonable since the fluctuators are generated
randomly from the distribution $1/f^{\alpha}$, and therefore for
higher values of $N_{f}$ there is a larger number of fluctuators
with low switching rates, and this situation keeps almost constant
the value of $q(t)$ for longer times, thus inducing a weaker decohering
effect on the system. 

We notice then that QC are destroyed a bit faster by brown noise,
and this is due to the fact that the distribution $1/f^{2}$ produces
a higher number of low switching-rate fluctuators with respect to
the $1/f$ distribution, therefore the system is more subject to back-action
from the environment, as we observed for RTN in the strong coupling
regime (see the previous section \ref{sub:BOS-Single-RTN-fluctuator}).
This is the reason for which the first decay of QC is faster for brown
noise (see for comparison the insets of Figs. \ref{fig:ED_Bos_RTN_nmk}
and \ref{fig:ED_Bos_RTN_mk}), but then we observe also a marked revival,
which is absent in the pink noise case. However, the difference in
the decay times is not so pronounced, due to the fact that in both
noise distributions there is also the effect of the fast-switching
fluctuators (which was absent in the RTN scenario for the strong coupling
regime), therefore the evolution of QC shows a hybrid character between
strong and weak coupling. With respect to the 2-qubit case, described
in Ref. \cite{Benedetti2013}, the effects of pink noise at $v=0$
are very similar, while for brown noise the dynamics of QC at $v=0$
are quite different, since in our system $1/f^{2}$ noise does not
give rise to entanglement sudden deaths and sudden births, as it happens
instead for qubits. Here indeed we observe the saturation of correlations
at long times, for the brown noise scenario (see Fig. \ref{fig:ED_BOS_PinkBrown_with_v}).

\begin{figure*}[!t]
\includegraphics[scale=0.45]{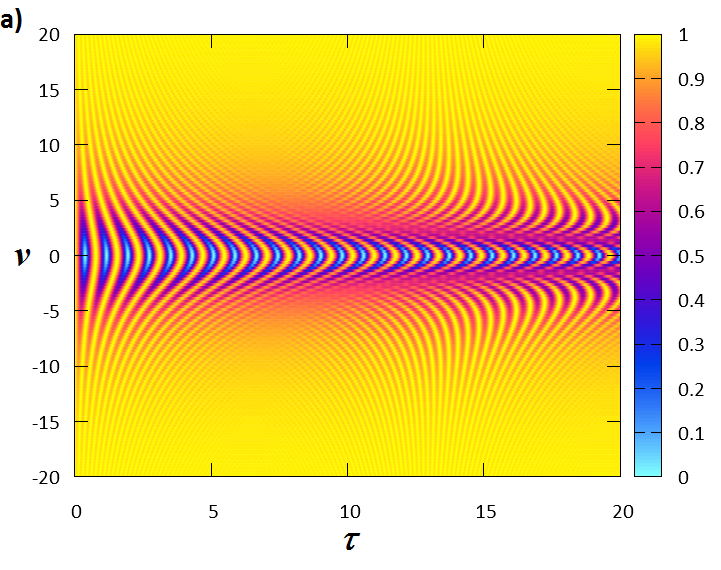}\hfill \includegraphics[scale=0.45]{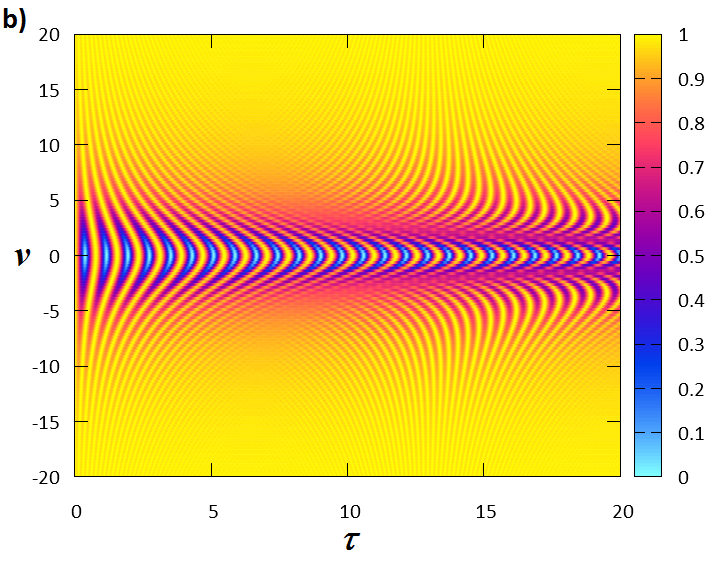}

\caption{Entanglement of particles and Discord of particles for the initial
state $\left|\Psi_{F}\right\rangle $. \textbf{a)} Entanglement $E_{P}$
\textbf{b)} Discord $D_{P}$.}
\label{fig:ED_fer_noiseless_Psi}

\begin{minipage}[t]{.49\textwidth}
\begin{centering}

\includegraphics[scale=0.4]{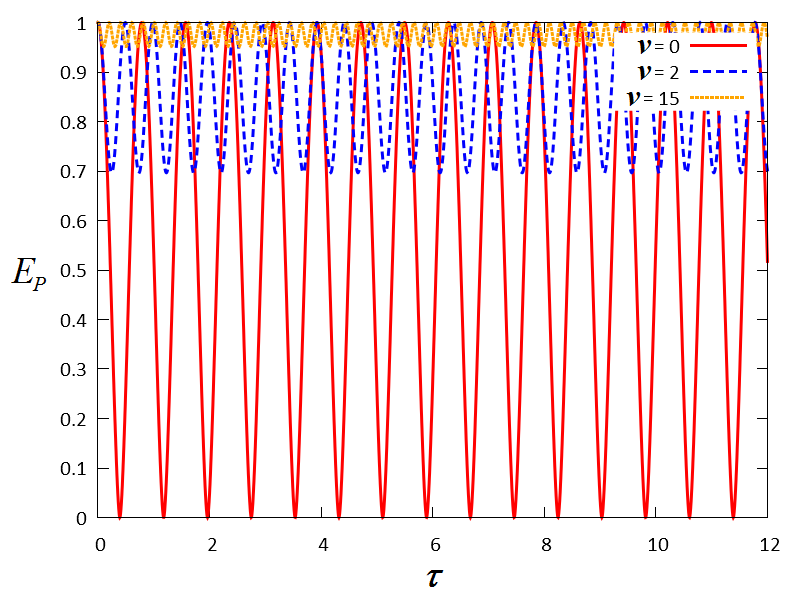}

\caption{Entanglement of particles for the initial state $\left|\Psi_{F}\right\rangle $
at different relative strength $v$ of interactions.}
\label{fig:E_fer_noiseless_Psi_oscill}

\end{centering}
\end{minipage}
\hfill
\begin{minipage}[t]{.49\textwidth}
\begin{centering}

\includegraphics[scale=0.45]{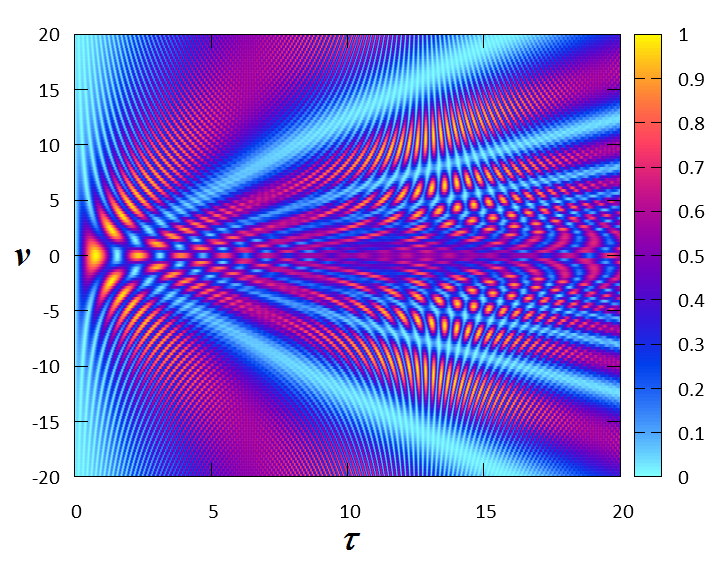}

\caption{Entanglement of particles $E_{P}$ for the initial state $\left|\Xi_{F}\right\rangle $
(see text for definition). The results for Discord of particles $D_{P}$
are not reported since they are identical.}
\label{fig:E_fer_noiseless_Xi}

\end{centering}
\end{minipage}
\end{figure*}

\section{Numerical simulations: Fermi-Hubbard model\label{sec:Numerical-simulations:-Fermi-Hubb}}

\subsection{Noiseless system\label{sub:FER-Noiseless-system}}

Here we describe the results of the numerical simulations for the
noiseless Fermi-dimer system. We start with the maximally entangled
state $\left|\Psi_{F}\right\rangle $ of Eq. \eqref{eq:Psi_F}, and
the evolution of the correlations are practically identical for Entanglement
$E_{P}$ (Fig. \ref{fig:ED_fer_noiseless_Psi} a) and Quantum Discord
$D_{P}$ (Fig. \ref{fig:ED_fer_noiseless_Psi} b), as we saw for the
bosonic case. Also in this case the behavior of the system is perfectly
symmetrical for positive and negative $v$.

To compare the situation with the boson case, we see that for $|v|>5$
the entanglement approaches a sort of limit behavior ($|v|\gtrsim15$),
with small oscillations around its maximum value (see Fig. \ref{fig:E_fer_noiseless_Psi_oscill}).
Therefore we deduce that the effect of interactions in preserving
the correlations is stronger for the fermionic system.

Due to the formal analogies between Eq. \eqref{eq:BosHubbHam} and
\eqref{eq:Fermi_Hubb_Hamilt}, we wanted to check whether the Hamiltonian
of Eq. \eqref{eq:Fermi_Hubb_Hamilt} allows to create a fermionic
state with behavior analogous to $\left|\Xi_{B}\right\rangle $. Therefore,
we initialized the system in the state $\left|\Xi_{F}\right\rangle =\frac{1}{\sqrt{2}}(c_{L\uparrow}^{\dag}+c_{R\uparrow}^{\dag})c_{R\downarrow}^{\dag}|0\rangle$,
which is the only possibility left%
\footnote{Except for symmetrical states that are equivalent to $|\Xi_{F}{\rangle}$
– within a relabeling of spins and/or sites – and a relative phase
between the two terms. Notice that $|\Upsilon_{F}{\rangle}=\frac{1}{\sqrt{2}}(c_{L\uparrow}^{\dag}c_{R\downarrow}^{\dag}+c_{R\uparrow}^{\dag}c_{L\downarrow}^{\dag})|0{\rangle}$
is equivalent to $|\Psi_{F}{\rangle}$ in terms of entanglement.%
} for having one particle in each partition of the system (notice that,
during the evolution, the particles cannot change their spin). Even
in this case, however, the evolution is symmetrical with respect to
the sign of $v$ (see Fig. \ref{fig:E_fer_noiseless_Xi}): this is
a consequence of anti-symmetrization and/or inhibition of spin-flip,
whose effect is to suppress many of the allowed transitions for the
bosonic system%
\footnote{Notice also that the \emph{Pauli exclusion principle} here is given
automatically by spin-flip suppression, therefore states like $|j,j{\rangle}$
would have remained unoccupied even if the particles initialized in
the state $|\Psi_{F}{\rangle}$ were bosons.%
}.

\begin{figure*}[!tp]
\includegraphics[scale=0.37]{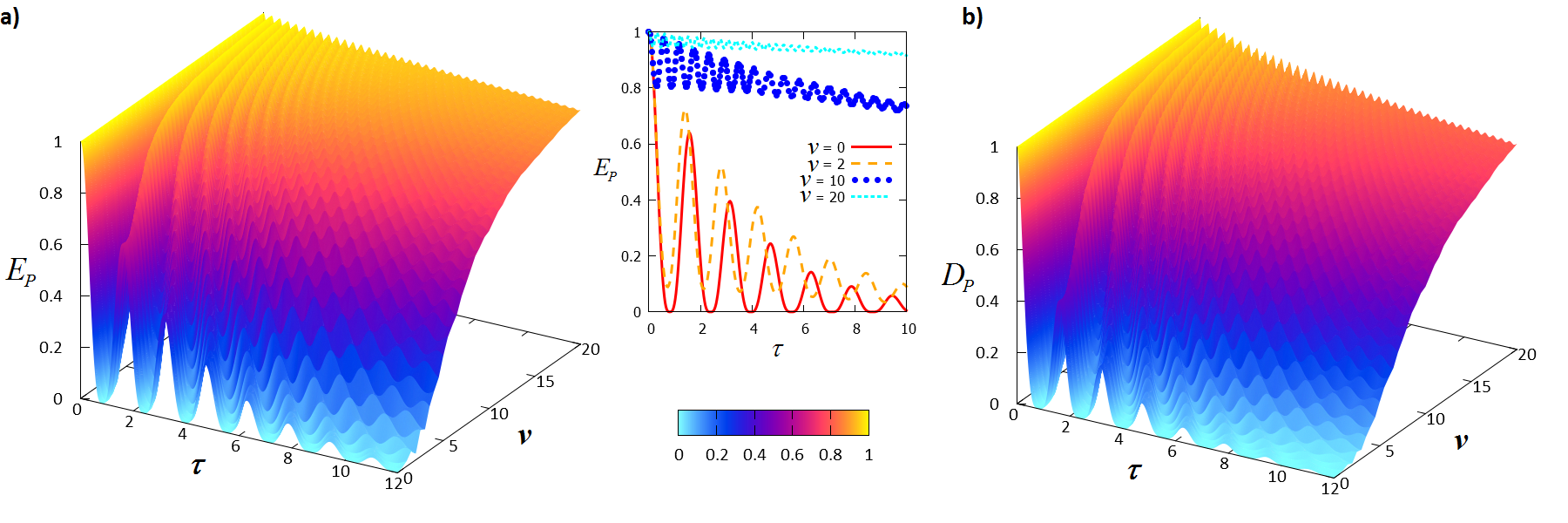}

\caption{Entanglement $E_{P}$ (a) and Discord of particles $D_{P}$ (b)for
$\left|\Psi_{F}\right\rangle $ with a single RTN fluctuator in the
strong coupling regime ($\gamma_{0}\tau_{s}=0.1$). \textbf{Inset:}
$E_{P}$ for different values of the interaction strength $v$.}
\label{fig:ED_Fer_RTN_nmk}
\end{figure*}

\begin{figure*}[!tp]
\includegraphics[scale=0.37]{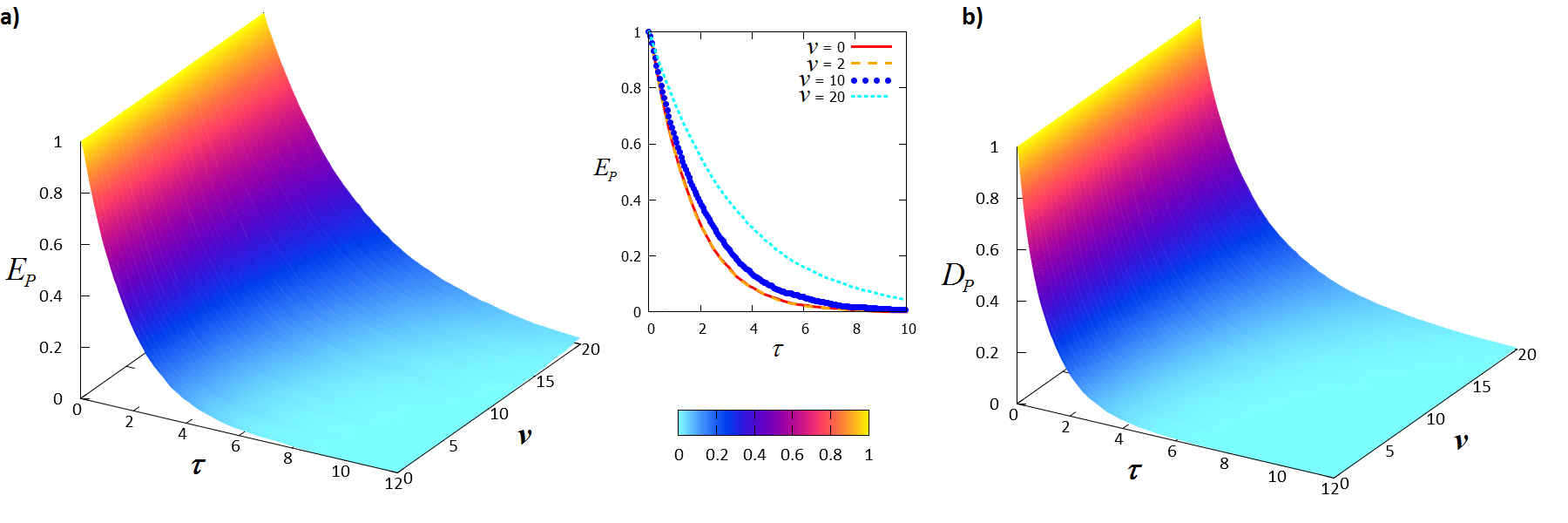}

\caption{Entanglement $E_{P}$ (a) and Discord of particles $D_{P}$ (b) for
$\left|\Psi_{F}\right\rangle $ with a single RTN fluctuator in the
weak coupling regime ($\gamma_{0}\tau_{s}=10$). \textbf{Inset:} $E_{P}$
for different values of the interaction strength $v$.}
\label{fig:ED_Fer_RTN_mk}
\end{figure*}

It is worth noting that the suppression of spin-flip transitions implies
that at any time each subsystem contains exactly one particle ($P_{1,1}=1$),
and therefore $E_{P}=\mathcal{E}(\rho_{1,1})$, $D_{P}=\mathcal{D}(\rho_{1,1})$.
This means that in general QC are stronger in the Fermi dimer than
in the bosonic system.

\subsection{Single RTN fluctuator\label{sub:FER-Single-RTN-fluctuator}}

Next, we study the effect of a single RTN fluctuator (with rate $\gamma_{0}$)
on the Fermi dimer, both in the strong ($\gamma_{0}\tau_{s}<1$) and
weak ($\gamma_{0}\tau_{s}>1$) coupling regime. Again the behavior
is perfectly symmetrical, so we show only the evolution of QC for
$v>0$. 

As can be seen from Figure \ref{fig:ED_Fer_RTN_nmk}, in the strong-coupling
regime both entanglement and discord are subject to a series of oscillations.
When $v=0$, we clearly observe sudden deaths and revivals, but when
$v>0$ the protective effect of $V$ prevents QC from going to zero
for a longer time, and at higher $v$ the effect is stronger, so that
decoherence is significantly reduced. As in the bosonic case, the
decay of discord is faster than that of entanglement.

For what concerns the strong-coupling regime – that is represented
in \ref{fig:ED_Fer_RTN_mk} – the oscillatory behavior is absent:
QC show a monotonic decay, which is again faster for quantum discord,
and the protective effect of interactions is still present but almost
negligible with respect to the previous case (see Figure \ref{fig:ED_Fer_RTN_mk},
Inset). We conclude that the protective action of interactions over
QC is effective only for low frequency noises and, in this condition,
it is much more effective than for bosons.

Both regimes are very similar to those observed in the literature
for quantum walks of classical (distinguishable) particles, subject
to a dynamical noise respectively in non-Markovian and Markovian regime.

\begin{figure*}
\includegraphics[scale=0.37]{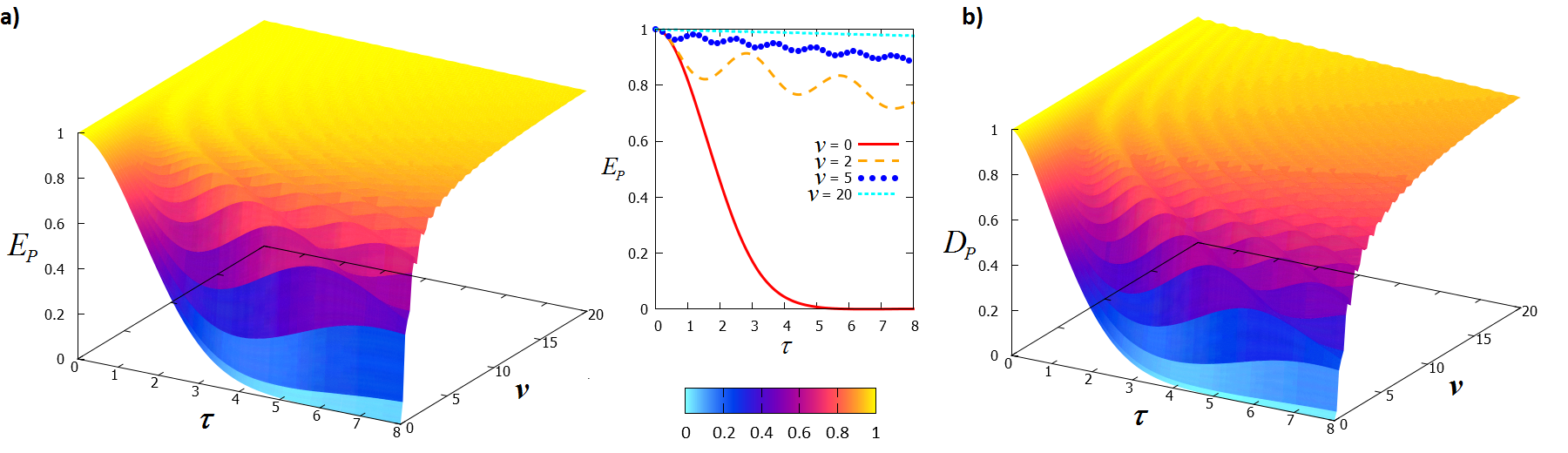}

\caption{Entanglement $E_{P}$ (a) and Discord of particles $D_{P}$ (b) for
$1/f$ (pink) noise, with $N_{f}=20$ fluctuators. \textbf{Inset}:
Evolution of $E_{P}$ for different strengths $v$ of the interactions.}
\label{fig:ED_FER_PinkNoise}
\end{figure*}

\begin{figure*}[t]
\includegraphics[scale=0.37]{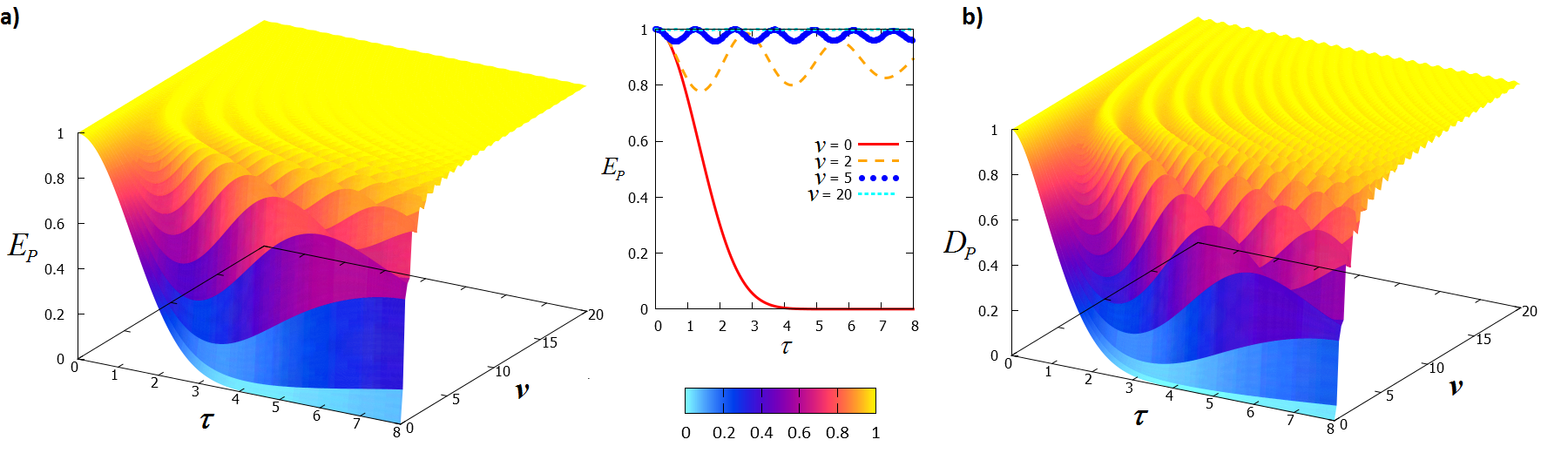}

\caption{Entanglement $E_{P}$ (a) and Discord of particles $D_{P}$ (b) for
$1/f^{2}$ (brown) noise, with $N_{f}=20$ fluctuators. \textbf{Inset}:
Evolution of $E_{P}$ for different strengths $v$ of the interactions.}
\label{fig:ED_FER_BrownNoise}
\end{figure*}

\begin{figure}[!bh]
\includegraphics[scale=0.4]{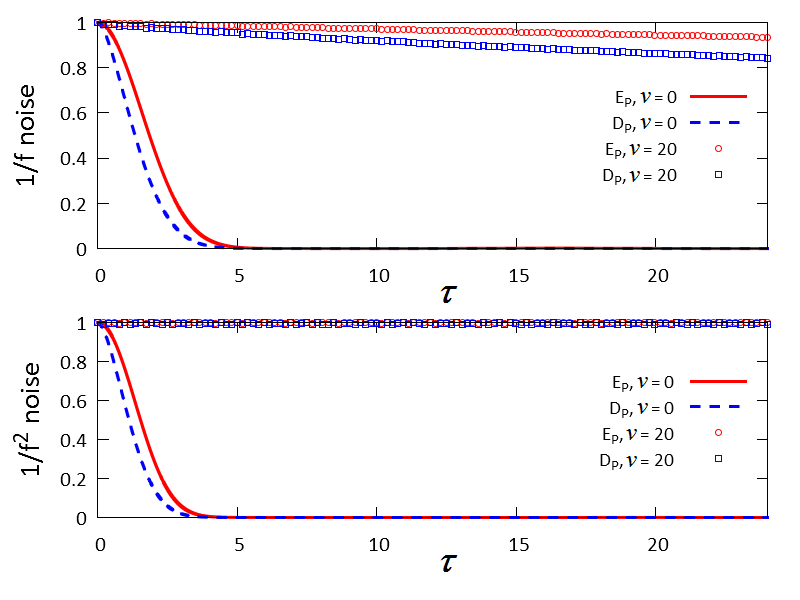} 

\caption{Entanglement $E_{P}$ (red) and Discord $D_{P}$ (blue) for $1/f$
(top panel) and $1/f^{2}$ (bottom panel) noise, with $N_{f}=20$
fluctuators, at different values of the interaction strength $v$
and for long time scales.}
\label{fig:ED_FER_PinkBrown_with_v}
\end{figure}

\subsection{Colored noises\label{sub:FER-Colored-noises}}

Moving to the colored noise scenario, we introduce again a perturbation
in the hopping amplitudes of electrons due to a collection of $N_{f}$
bistable fluctuators, whose switching frequencies $\gamma$ are distributed
with the power law of pink or brown noise and are randomly generated
in the interval $\gamma\in|T|\cdot[1.25\cdot10^{-4},1.25\cdot10^{2}]$.

We observe again a situation which is quite different with respect
to the bosonic case, since the effect of interactions over QC is very
strong in this scenario, both for pink and brown noise. In both cases,
the decay of correlations is fast for $v=0$ and very slow for $v>0$
(see Figs. \ref{fig:ED_FER_PinkNoise}, \ref{fig:ED_FER_BrownNoise}
and \ref{fig:ED_FER_PinkBrown_with_v}), practically negligible at
high $v$ in the brown noise scenario (see Fig. \ref{fig:ED_FER_PinkBrown_with_v}).
Again, we observe that discord has the same qualitative behavior of
entanglement but a faster decay. Moreover, as for bosons, the decay
of QC at $v=0$ is slightly faster in the brown noise scenario, but
the differences between the two noisy environments are very small.
Both for pink and brown noise, the decay of QC for $v>0$ is oscillatory
(see Insets of Figs. \ref{fig:ED_FER_PinkNoise} and \ref{fig:ED_FER_BrownNoise})
– a difference with respect to the bosonic case – but the amplitudes
get smaller and the oscillations are faster at higher $v$. 

For what concerns the number of fluctuators $N_{f}$, we observe the
same effects seen for bosons, namely a slower decay of QC for a higher
number of fluctuators, which is again due to the renormalization of
noise.


\section{Conclusions\label{sec:Conclusions}}

In this paper we propose an expression for the \emph{quantum discord}
$D_{P}$ of a couple of indistinguishable particles through the generalization
of the notion of \emph{entanglement of (indistinguishable) particles
$E_{P}$} introduced by \emph{Wiseman and Vaccaro}\cite{Wiseman2003}.
We used this quantifier of quantum correlations for studying two 1D
lattice model systems, both described by the Hubbard Hamiltonian,
in which a couple of interacting bosons or fermions tunnels between
the sites of a chain subject to classical environmental noise (RTN,
pink $1/f$, brown $1/f^{2}$). We confronted the results for $D_{P}$
with the corresponding values of $E_{P}$, showing that in all the
studied cases there are no marked qualitative differences between
the two quantifiers: the peculiarity of this behavior (which is expected
for pure and not for mixed states) in our opinion is due to the dynamics
of the noise (since it was observed also for qubits subject to classical
environments \cite{Benedetti2012,Benedetti2013}), but at the same
time it is a strong evidence of the reliability of the proposed measure
$D_{P}$ as a quantifier for quantum correlations of indistinguishable
particles. In detail, for pure states $D_{P}$ always coincides with
entanglement of formation, while for noisy systems $D_{P}$ is generally
lower than $E_{P}$: a situation which is not common but it has already
been observed in the literature, both in bipartite and multipartite
systems of qubits\cite{Luo2008a,beggi2015} and in particular under
the effect of classical noise \cite{Benedetti2012,Benedetti2013}.
On the other hand, in one of the studied cases we observed that the
entanglement goes to zero while the quantum discord does not vanish,
as it is usually observed in the literature. We expect that a different
behavior of $E_{P}$ and $D_{P}$ will be more evident when the model
is changed (adding e.g. external fields or more complex interactions)
or when the proposed quantifier will be used to describe other kinds
of systems composed by identical particles: both cases, however, are
beyond the scope of the present work, but can be considered as promising
developments of this research.

As far as the dynamic of the systems is concerned, we observed – both
in the bosonic Hubbard plaquette and in the fermionic Hubbard dimer
– that in general the effect of the on-site interactions $V$ over
quantum correlations does not depend on its sign: a symmetry already
described in the literature\cite{Lahini2012,schneider2010} and attributed
to the invariance of the bandstructure with respect to $\mathrm{sgn}(V)$.
However, for the bosonic system, an exception was found in the state
$\left|\Xi_{B}\right\rangle =\frac{1}{\sqrt{2}}(b_{4}^{\dag}b_{1}^{\dag}+b_{4}^{\dag}b_{2}^{\dag})|0\rangle$,
where the switching from repulsive to attractive interactions modified
deeply the evolution of both entanglement and discord. This phenomenon
requires a further analysis, since it has never been observed before,
and it can offer new interesting possibilities for the controlled
manipulation of quantum correlations. 

We also notice that the interactions $V$ are able to inhibit the
decoherence of the system by preserving QC (i.e., both entanglement
and discord): the stronger is $V$, the higher is the degree of protection,
and this effect is much more marked in the fermionic system. To a
further level of detail, we notice that the protective action of the
on-site interaction $V$ over correlations is much more effective
for low-frequency noises than for high-frequency ones. This effect
is clearly visible for RNT noise in the weak and strong coupling regimes,
but also partially for brown and pink noises, since the former scenario
is dominated by low-frequency fluctuators much more than the latter,
and therefore it shows a higher ``effective back-action'' of the
environment over the system. This is the reason why we observe a higher
resistance to decoherence (fermions) and a revival of quantum correlations
(bosons) for the systems subject to brown noise, while these phenomena
are weaker or absent in the pink noise scenario.

The main difference between the bosonic and the fermionic system is
that the latter is able to preserve QC in a more efficient way than
the former, even with a lower value of the relative interaction strength
$v$. On the other hand, these discrepancies can be attributed to
some basic distinctive elements, such as the differences in geometry
of the two models (two vs. four sites) and the lower number of allowed
transitions in the fermionic system, due both to spin-flip suppression
and anti-symmetrization.

These results evidence many connections with those obtained in systems
of quantum bits or classical particles subject to RTN or $1/f^{\alpha}$
noises\cite{Benedetti2012,Bordone2012,Benedetti2013}, thus showing
that some previous results concerning model system of qubits can act
as a guideline to interpret experimental observations on quantum random
walk, also suggesting new possible directions of investigation.

Further perspective of research on quantum correlations could be the
extension of the interaction to particles occupying neighboring sites\cite{Qin2014}
and possibly the introduction of a constant external field: two aspects
which could be used to engineer or preserve quantum correlations at
a deeper and more efficient level. Also the effects of static noise
(resulting in lattice disorder) could be explored in the case of identical
particles – due to their role in phenomena like Anderson localization
or the transition from quantum to classical random walks\citep{Amir2009, *Lahini2010, *Thompson2010, *benedetti2015}–
in order to explore differences, if present, with the distinguishable
particle case\cite{Bordone2012}.

\subsection*{Acknowledgments}

This work has been supported by UniMORE through FAR2014.

\appendix

\section{Properties of Quantum Discord\label{sec:Appendix_QD}}

Quantum discord $\mathcal{D}_{B}(\rho_{AB})$ gives the amount of
correlations between the subsystem $A$ and $B$ that are destroyed
by a measurement on $B$, which can be interpreted as a measure of
quantum correlations between the two subsystems. The properties required
for a good quantifier of \emph{quantum discord} are the following
ones\cite{Henderson2001,Modi2012}:

\begin{enumerate}[(i).]

\item discord is non-negative;

\item discord is not symmetric, i.e. in general $\mathcal{D}_{B}(\rho_{AB})=S(\rho_{1,1}^{B})+S(\rho_{1,1}^{A}|\{\Pi_{k}^{B}\})-S(\rho_{1,1})\ne\mathcal{D}_{A}(\rho_{AB})$
depends upon the subsystem on which the measurements $\{\Pi_{k}\}$
are performed;

\item discord is invariant under local-unitary transformations, i.e.
$\mathcal{D}_{S}(\rho_{AB})=\mathcal{D}_{S}((U_{B}^{\dagger}\otimes U_{A}^{\dagger})\rho_{AB}(U_{A}\otimes U_{B}))$
for any couple of unitary transformations $U_{A},\, U_{B}$;

\item discord is non-decreasing under local operations;

\item discord is a monotone of entanglement for pure states, i.e.
$\mathcal{D}_{A}(\rho_{AB})=S(\rho_{A})=S(\rho_{B})$;

\item discord vanishes if the state $\rho_{AB}$ is classical-quantum
with respect to the measured subparty $A$, i.e. $\rho_{AB}=\Sigma_{i}p_{i}\Pi_{i}^{A}\otimes\rho_{i}^{B}$
(we can consider classical states as a subclass of classical-quantum
states);

\item discord is bounded from above, as $\mathcal{D}_{B}(\rho_{AB})\le S(\rho_{B})$.

\end{enumerate}

As we will show briefly, our quantifier for discord of particles $D_{P}$
possesses all the above properties, and therefore it can be considered
as a measure for quantum correlations of identical particles.

For a system of $N$ indistinguishable particles, shared among two
subparties $A$ and $B$, $D_{P}=D_{P}^{(S)}$ assumes the following
form
\begin{equation}
D_{P}^{(S)}=\sum_{k=0}^{N}P_{k,N-k}\mathcal{D}_{S}(\rho_{k,N-k}),
\end{equation}
where $S$ is the measured subsystem and
\begin{equation}
\rho_{k,N-k}=\Pi_{k,N-k}\rho\Pi_{k,N-k}
\end{equation}
is the projection of the system state $\rho$ over the subspace in
which $A$ possesses exactly $k$ particles and $B$ the remaining
$N-k$ ones, while $P_{k,N-k}=\mathrm{Tr}_{AB}[\rho_{k,N-k}]$ is
the corresponding probability. In detail, if $A$ controls the modes
$\{n_{A1},n_{A2},...\}$ and $B$ controls the modes $\{n_{B1},n_{B2},...\}$,
we can write the projector $\Pi_{k,N-k}$ as:
\begin{align}
\Pi_{k,N-k} & =\sum_{\substack{\Sigma_{i}n_{Ai}=k\\
\Sigma_{i}n_{Bi}=N-k
}
}\left|\{n_{Ai}\}\{n_{Bi}\}\right\rangle \left\langle \{n{}_{Ai}\}\{n{}_{Bi}\}\right|,\nonumber \\
 & =\sum_{\Sigma_{i}n_{Ai}=k}\left|\{n_{Ai}\}\right\rangle \left\langle \{n{}_{Ai}\}\right|\otimes\\
 & \sum_{\Sigma_{i}n_{Bi}=N-k}\left|\{n_{Bi}\}\right\rangle \left\langle \{n{}_{Bi}\}\right|,\nonumber 
\end{align}
where we used the property $\forall\left|\{n_{Ai}\}\{n_{Bi}\}\right\rangle \quad\Sigma_{i}n_{Ai}=k\iff\Sigma_{i}n_{Bi}=N-k$.
Of course, the sum of all the projectors gives the identity:
\begin{equation}
\sum_{k=0}^{N}\Pi_{k,N-k}=I.
\end{equation}

Now the reduced matrices for the two subsystems can be written, e.g.,
as:
\begin{align}
\rho_{k,N-k}^{B} & =\frac{\mathrm{Tr}_{A}[\rho_{k,N-k}]}{\mathrm{Tr}_{AB}[\rho_{k,N-k}]}\nonumber \\
 & =\frac{\mathrm{Tr}_{A}[\rho_{k,N-k}]}{P_{k,N-k}}\\
 & =\frac{1}{P_{k,N-k}}\sum_{\{n_{Ai}\}}\left\langle \{n{}_{Ai}\}\right|\rho_{k,N-k}\left|\{n_{Ai}\}\right\rangle \nonumber \\
 & =\frac{1}{P_{k,N-k}}\sum_{\Sigma_{i}n_{Ai}=k}\left\langle \{n{}_{Ai}\}\right|\rho_{AB}\left|\{n_{Ai}\}\right\rangle .\nonumber 
\end{align}

Therefore, it is easy to see that:
\begin{align}
\rho_{B} & =\mathrm{Tr}_{A}[\rho_{AB}]\nonumber \\
 & =\sum_{\{n_{Ai}\}}\left\langle \{n{}_{Ai}\}\right|\rho_{AB}\left|\{n_{Ai}\}\right\rangle \\
 & =\sum_{k}\sum_{\Sigma_{i}n_{Ai}=k}\left\langle \{n{}_{Ai}\}\right|\rho_{AB}\left|\{n_{Ai}\}\right\rangle \nonumber \\
 & =\sum_{k}P_{k,N-k}\rho_{k,N-k}^{B}.\nonumber 
\end{align}

Property (i) is immediately verified, since $D_{P}$ is a convex combination
of non-negative quantities. The same discussion holds for property
(ii), since $D_{P}^{(S)}$ is also a convex combination of non-symmetrical
quantifiers with respect to $S$.

Let us recall now a property of the local operations $L_{A},\, L_{B}$
(acting on $A$ or $B$) stated in Ref~\cite{Wiseman2003}: they
do not change the local number of particles, therefore they commute
with the operation of measuring the local number of particles and
conserve the probability $P_{k,N-k}$:
\begin{align}
\Pi_{k,N-k}L_{A}\rho L_{A}^{\dagger}\Pi_{k,N-k} & =L_{A}\Pi_{k,N-k}\rho\Pi_{k,N-k}L_{A}^{\dagger}
\end{align}
\begin{align}
P_{k,N-k} & =\mathrm{Tr}_{AB}[\rho_{k,N-k}]\nonumber \\
 & =\mathrm{Tr}_{AB}[L_{A}\rho_{k,N-k}L_{A}^{\dagger}]
\end{align}

Using this property, it is immediate to prove properties (iii) and
(iv) since $D_{P}$ is a convex combination (with constant coefficients
$P_{k,N-k}$) of non decreasing quantities $\mathcal{D}_{S}(L_{A}\rho_{k,N-k}L_{A}^{\dagger})\ge\mathcal{D}_{S}(\rho_{k,N-k})$,
which are also invariant if $L_{A}$ is unitary.

If we consider a pure state, recalling that the projective operations
$\Pi_{k,N-k}$ do not change its nature, we use the properties of
$\mathcal{D}_{A}(\rho_{AB})$ and we get:
\begin{align}
D_{P}^{(A)} & =\sum_{k=0}^{N}P_{k,N-k}S(\mathrm{Tr}_{B}[\rho_{k,N-k}])\nonumber \\
 & =\sum_{k=0}^{N}P_{k,N-k}S(\rho_{k,N-k}^{A}),
\end{align}
which proves property (v) (notice that for these states $D_{P}=E_{P}$
if we use the von Neumann entropy as a measure of entanglement).

Then, when considering a classical-quantum state $\rho_{AB}=\Sigma_{i}p_{i}\Pi_{i}^{A}\otimes\rho_{i}^{B}$,
its projection $\Pi_{k,N-k}\rho_{AB}\Pi_{k,N-k}$ is still a classical-quantum
state, and therefore $D_{P}^{(A)}$ results in a linear combination
of vanishing quantities, giving thus property (vi). 

Finally, it is easy to prove that $D_{P}$ is bounded from above (vii)
since it is a convex combination of quantities that are bounded from
above:
\begin{align}
D_{P}^{(B)} & =\sum_{k=0}^{N}P_{k,N-k}\mathcal{D}_{B}(\rho_{k,N-k})\nonumber \\
 & \le\sum_{k=0}^{N}P_{k,N-k}S(\rho_{k,N-k}^{B})\le S(\rho_{B}),
\end{align}
where the last passage comes from the concavity property $S(\Sigma_{i}p_{i}\rho_{i})\ge\Sigma_{i}p_{i}S(\rho_{i})$
of Von Neumann entropy.

\section{Purity and Decoherence\label{sec:Appendix_Pur}}

\begin{figure*}[t]
\begin{minipage}[t]{.49\textwidth}
\begin{centering}

\begin{centering}
\includegraphics[scale=0.4]{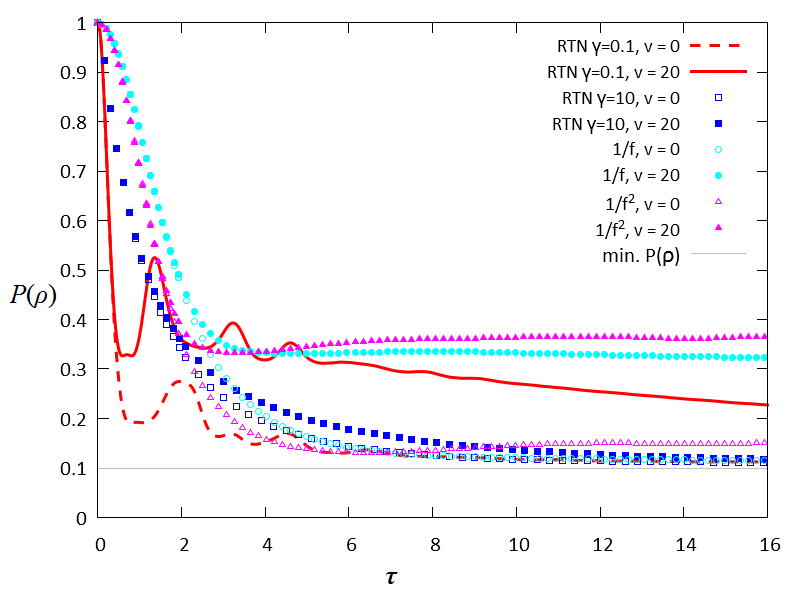}
\par\end{centering}

\caption{Purity of the state $\left|\Psi_{B}\right\rangle $ under different
classical noises and interaction strengths $v$}
\label{fig:Pur_bos}

\end{centering}
\end{minipage}
\hfill
\begin{minipage}[t]{.49\textwidth}
\begin{centering}

\begin{centering}
\includegraphics[scale=0.4]{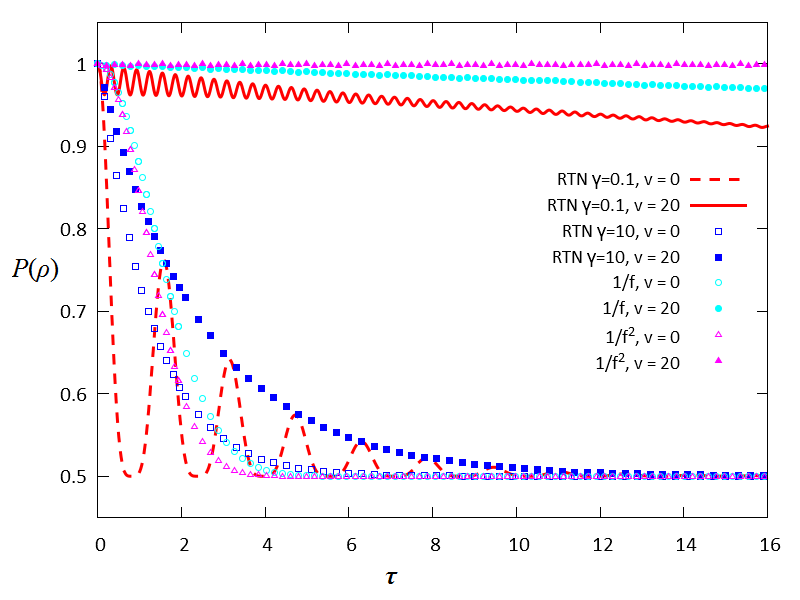}
\par\end{centering}

\caption{Purity of the state $\left|\Psi_{F}\right\rangle $ under different
classical noises and interaction strengths $v$}
\label{fig:Pur_fer}

\end{centering}
\end{minipage}
\end{figure*}

In all the analyzed cases, our \emph{discord of particles }is very
similar to entanglement of particles and, moreover, it is typically
a bit smaller than it. This behavior is uncommon, since usually discord
tends to be higher than entanglement even under external detrimental
agents (such as, e.g., high temperatures or strong magnetic fields\cite{Werlang2010}).
A typical condition in which there's no significant difference between
entanglement and discord is when the considered state is almost pure,
but this is definitely not the case for our system, whose states are
heavily mixed by noise. The mixing of the state can be quantified
with the purity,
\begin{equation}
P(\rho_{AB})=\mathrm{Tr}[\rho_{AB}^{2}]
\end{equation}
which is $1$ for a pure state and $1/d$ ($d=\mathrm{dim}(\rho_{AB})$)
for a maximally mixed state, but also through the \emph{decoherence},
which is quantified by the von Neumann entropy
\begin{equation}
S_{D}(\rho_{AB})=-\frac{1}{\ln(d)}\rho_{AB}\ln\rho_{AB}.\label{eq:Decoh}
\end{equation}
To be more precise, we should mention that the decoherence of this
system should not be interpreted as a measure of the entanglement
between the system and the environment, since our environment is not
included in the quantum evolution, but it is modeled in a classical
way. As a result, there isn't a true flow of information between the
system and the environment, but noise is still capable of mixing the
density matrix, thus determining a loss of coherence in $\rho_{AB}$
which is quantified by Eq. \eqref{eq:Decoh}. We computed both quantities
for our simulations, but in the following we report only purity for
reasons of clarity and brevity, since the two quantifiers agree qualitatively
in all considered cases.

As it can be seen in Fig. \ref{fig:Pur_bos}, the action of classical
noise over the bosonic system drives it towards a heavily mixed state.
Mixing is almost maximal for $V=0$ - where purity goes to $1/d=0.1$
and decoherence (not shown) goes to $1$ - while at high strengths
of the interaction the loss of coherence is limited, in agreement
with our observations on the role of $V$ in preserving quantum correlations.
A similar behaviour is observed for the fermionic system, shown in
Fig. \ref{fig:Pur_fer}, but here even at $V=0$ mixing is never maximal
(purity downfalls and saturates at $0.5$, which is larger than $1/d\simeq0.17$),
coherently with the fact that we observed a higher resistance of correlations
to noise in this kind of system. Moreover, we notice that at high
$V/J$ there is almost no mixing - as well as there is almost no loss
in quantum correlations - for slow RTN and $1/f^{\alpha}$ noises
(see Figs. \ref{fig:ED_Fer_RTN_nmk} and \ref{fig:ED_FER_PinkNoise}-\ref{fig:ED_FER_PinkBrown_with_v}
for comparison).

Starting the simulation with a state which is not maximally entangled
does not change the dynamics of correlations: e.g., the state $\left|\Psi'_{B}\right\rangle =\frac{1}{2}\left(b_{1}^{\dagger}b_{3}^{\dagger}+\sqrt{3}b_{2}^{\dagger}b_{4}^{\dagger}\right)\left|0\right\rangle $
still shows an evolution in which discord decays faster than entanglement
under the action of noise (even if, at larger times, they cross before
they go both to zero). 

Moreover, the same relationship between entanglement and discord (i.e.,
a very similar evolution in time and the hierarchy $D<E$), was observed
for quantum bits subject to classical RTN \cite{Benedetti2012} and
colored noise \cite{Benedetti2013}: we therefore conclude that this
behavior of entanglement and discord can be related to the peculiar
features of the noise generated by randomly switching bistable fluctuators
and is neither a consequence of the choice of the initial state nor
an effect of low mixing.


%

\end{document}